\documentclass{aa}  

\usepackage{graphicx}
\usepackage{txfonts}
\usepackage{xcolor}
\usepackage{cleveref}

\newcommand{\xidm}{\xi_{\rm{DM}}}

\newcommand{\de}{\ensuremath{\mathrm{d}}}

\newcommand{\Mpc}{\mathrm{Mpc}}

\newcommand{\ob}{^{\rm ob}}
\newcommand{\true}{^{\rm true}}

\newcommand*{\Msun}{\ensuremath{\mathrm{M_\odot}\,\,}}%
\newcommand{\sect}[1]{Sect.~#1}
\newcommand{\fig}[1]{Fig.~#1}
\newcommand{\eq}[1]{Eq.~#1}
\newcommand{\tb}[1]{Table~#1}

\newcommand{\figs}[1]{Figs.~#1}
\newcommand{\eqs}[1]{Eqs.~#1}

\defcitealias{Costanzi2019ab}{Co19}
\defcitealias{Finoguenov2019aa}{Fi19}
\defcitealias{Bocquet2019aa}{Bo19}
\defcitealias{Pillepich2018aa}{Pi18}
\defcitealias{Pacaud2018aa}{Pa18}
\defcitealias{Mantz2014aa}{Ma14}
\defcitealias{Zu2014aa}{Zu14}
\defcitealias{Zubeldia2019aa}{Zb19}
\defcitealias{Kirby2019aa}{Ki19}

\newcommand{\furl}[1]{\footnote{\url{#1\xspace}}}

\newcommand{\Om}{\Omega_{m_{0}}}

\newcommand{\Ode}{\Omega_{\rm{DE}_{0}}}

\newcommand{\normxs}[2]{%
\frac{1}{\sqrt{2\pi}#2}e^{{{-\frac{1}{2}\left({#1}\right)^2}\mathord{\left/{\vphantom{{-\left({#1}\right)^2}{#2^2}}}\right.\kern-\nulldelimiterspace}{#2^2}}}
}
\newcommand{\normxms}[3]{%
\frac{1}{{\sqrt{2\pi}#3}}\exp\left[{{-\frac{1}{2}\left({#1-#2}\right)^2}\mathord{\left/{\vphantom{{-\left({#1-#2}\right)^2}{#3^2}}}\right.\kern-\nulldelimiterspace}{#3^2}\right]}
}

\newcommand{\wpdm}{w_p^{DM}}

\newcommand{\xivec}{\pmb{\xi}}
\newcommand{\xidmvec}{\pmb{\xi}^\mathrm{DM}}
\newcommand{\pimax}{\pi_\mathrm{max}}
\newcommand{\pimin}{\pi_\mathrm{min}}

\begin{document}

   \title{Clustering of CODEX clusters}

   \subtitle{}

   \author{V.~Lindholm
          \inst{1,2}
          \and
          A.~Finoguenov\inst{1}
          \and J.~Comparat\inst{3}
          \and C.~C.~Kirkpatrick\inst{1,2}
          \and E.~Rykoff\inst{4}
          \and N.~Clerc\inst{5}
          \and C.~Collins\inst{6}
          \and S.~Damsted\inst{1}
          \and J.~Ider Chitham\inst{3}
          \and N.~Padilla\inst{7}
          }

   \institute{Department of Physics,
              Gustaf H{\"a}llstr{\"o}min katu 2 A,
              University of Helsnki, Helsinki, Finland
         \and
             Helsinki Institute of Physics,
             Gustaf H{\"a}llstr{\"o}min katu 2,
             University of Helsinki, Helsinki, Finland
        \and 
        Max-Planck institute for Extraterrestrial physics, Giessebachstr, Garching 85748, Germany
        \and
        Kavli Institute for Particle Astrophysics \& Cosmology, P. O. Box 2450, Stanford University, Stanford, CA 94305, USA; SLAC National Accelerator Laboratory, Menlo Park, CA 94025, USA
        \and
     IRAP, Universit{\'e} de Toulouse, CNRS, UPS, CNES, Toulouse, France
     \and
     Astrophyics Research Institute, Liverpool John Moores University, IC2, Liverpool Science Park, 146 Brownlow Hill, Liverpool L3 5RF, UK
     \and
     Instituto de Astrof{\'i}sica, Pontificia Universidad Cat{\'o}lica de Chile, Av. Vicuna Mackenna 4860, 782-0436 Macul, Santiago, Chile
             }

   \date{Received \today; accepted later}

 
  \abstract
      {The clustering of galaxy clusters
  links the spatial nonuniformity of dark matter halos to the growth
  of the primordial spectrum of perturbations. The amplitude of the
  clustering signal is widely used to estimate the halo mass of
  astrophysical objects. The advent of cluster mass calibrations
  enables using clustering in cosmological studies.}
   {We analyze the autocorrelation function of a large contiguous sample of galaxy clusters, the Constrain Dark Energy with X-ray (CODEX) sample, in which we take particular care of cluster definition. These clusters were X-ray selected using the ROentgen SATellite (ROSAT) All-Sky Survey (RASS) and then identified as galaxy clusters using the code redMaPPer run on the photometry of the Sloan Digital Sky Survey (SDSS). We develop methods for precisely accounting for the sample selection effects on the clustering and demonstrate their robustness using numerical simulations.}
   {Using the clean CODEX sample, which was obtained by applying a redshift-dependent richness selection, we computed the two-point autocorrelation function of galaxy clusters in the $0.1<z<0.3$ and $0.3<z<0.5$ redshift bins. We compared the bias in the measured correlation function with values obtained in numerical simulations using a similar cluster mass range.}
   {By fitting a power law, we measured a correlation length $r_0=18.7 \pm 1.1$ and slope $\gamma=1.98 \pm 0.14$  for the correlation function in the full redshift range. By fixing the other cosmological parameters to their nine-year Wilkinson Microwave Anisotropy Probe (WMAP) values, we reproduced the observed shape of the correlation function under the following cosmological conditions: $\Om=0.22^{+0.04}_{-0.03}$ and $S_8=\sigma_8 (\Om /0.3)^{0.5}=0.85^{+0.10}_{-0.08}$ with estimated additional systematic errors of $\sigma_{\Om} = 0.02$ and $\sigma_{S_8} = 0.20$. We illustrate the complementarity of clustering constraints by combining them with CODEX cosmological constraints based on the X-ray luminosity function, deriving  $\Om = 0.25 \pm 0.01$ and $\sigma_8 = 0.81^{+0.01}_{-0.02}$ with an estimated additional systematic error of $\sigma_{\Om} = 0.07$ and $\sigma_{\sigma_8} = 0.04$. The mass calibration and statistical quality of the mass tracers are the dominant source of uncertainty.}
   {}


   \maketitle
%

\section{Introduction}

Galaxy clusters are the largest gravitationally bound objects in the
Universe. A common assumption is that they completely and uniquely
trace the population of the high-mass dark matter halos. Thus galaxy
clusters form a powerful probe for the large-scale structure of the
Universe. This power is further enhanced by the fact that several of
the galaxy cluster properties, such as the abundance as a function of
total mass and spatial clustering, can be predicted based on a
cosmological model. A particularly important aspect of the clustering
of clusters is that they are biased tracers of the underlying matter
distribution, meaning that their clustering amplitude is higher than
that of the full distribution of matter. Moreover, the amount of
enhancement grows with increasing cluster masses. By studying the
relationship between cluster masses and bias, we can gain further
insights into the cosmology (see, e.g., \citealp{mo_white96} and
\citealp{sheth_tormen99})

Because the clustering amplitude of dark matter halos of a given mass
is sensitive to the underlying cosmology, the application of the
clustering theory to galaxy clusters is theoretically highly
motivated. Moreover, galaxy clusters exhibit scaling relations between
their baryonic properties and the total mass of their hosting
halos. These properties include the cluster X-ray luminosity and
richness (i.e., the number of galaxies belonging to the
cluster). Using these observable quantities as mass proxies on the one
hand and the connection between cluster masses and their bias on the
other, we can make inference on the cosmological model. Most attempts
to follow this route have resulted in cosmological parameters that
disagree with the constraints obtained using the number counts of the
same sample \citep{schuecker02}, however, or lead to strongly
disagreeing scaling relations \citep{jimeno17}. Remarkably,
\cite{allevato12} obtained a precise agreement between the modeling of
the clustering based on the weak-lensing mass calibration and the
measured clustering amplitudes of galaxy groups by taking the
definition of the object and the effects of sample variance into
account.

The goal of this paper is to measure the clustering of galaxy clusters
detected in the galaxy cluster survey called Constrain Dark Energy
with X-ray (CODEX) \citep{codex} by computing their auto two-point
correlation function. A brief description of the CODEX catalog is
given in \sect{\ref{sec:codex_survey}}. Furthermore, we aim at
predicting the excess clustering, quantified by the bias factor, based
on the cluster masses and a cosmological prediction for the total
matter distribution. The details of this procedure are presented in
\sect{\ref{sec:methods}}, which is a continuation of the developments
presented in \citet{allevato11} and \citet{allevato12}. In
\sect{\ref{sec:validation}} we apply this pipeline to a simulated dark
matter halo catalog to ensure that the predicted and measured
clustering amplitudes match. In \sect{\ref{sec:codex_clustering}} we
present the clustering measurements for the CODEX cluster
catalog. Finally, in \sect{\ref{sec:cosmology}} we use the measured
and predicted clustering amplitudes to constrain two parameters within
the spatially flat $\Lambda$CDM cosmology: the matter density
parameter $\Om$ , and the power spectrum normalization
$\sigma_8$. Similar analyses have been performed, for example, by
\citet{marulli18} using data from another X-ray selected survey, the
XXL survey \citep{xxl}, and by \citet{mana13} for a cluster sample
selected using the maxBCG red-sequence method from Sloan Digital Sky
Survey (SDSS) photometric data \citep{koester07}. A key improvement
with the CODEX catalog compared to the first analysis is the
significantly larger sample size (1892 clusters in our analysis here
compared to 187 clusters), and compared to the second analysis, we
gain an improvement through the X-ray selection.


\section{CODEX galaxy cluster survey and its modeling}
    \label{sec:codex_survey}
    
The CODEX galaxy cluster survey is constructed by performing the
detection of faint X-ray sources in the ROentgen SATellite (ROSAT)
All-Sky Survey (RASS) and a subsequent identification of those sources
using the redMaPPer algorithm \citep{redmapper} applied to the SDSS
photometry inside the 10,000 square degree area of the Baryon
Oscillation Spectroscopic Survey (BOSS) footprint. A detailed
description of the survey and the catalog is presented in
\cite{codex}. Spectroscopic identification of the sample has been a
target of the SPectroscopic IDentification of eROSITA Sources
(SPIDERS) program of SDSS-IV, with first results presented in
\citet{clerc16} and \citet{clerc20}. These first results confirmed the
high quality of the redMaPPer redshifts and the virialized nature of
CODEX clusters.
      
To perform the clustering analysis, we selected the clean CODEX
catalog by applying a richness cut
      \begin{equation}
          \exp(\lambda) > 22(z/0.15)^{0.8},
      \end{equation}
      where z denotes the redshift of the cluster and
      $\lambda\equiv\ln (\mathrm{SDSS\; Richness})$ \citep[defined at
        the optical peak, with a detailed description provided
        in][]{redmapper}. We describe the effect of this cleaning as a
      $P^\mathrm{RASS}(I | \lambda, z)$ term in the modeling. Because
      we use multivariate log-normal distributions throughout this
      paper, we conveniently define the quantities $r_c\equiv\ln
      (R_c)$ (core radius of the X-ray surface brightness),
      $l\equiv\ln (L_x)$ (rest-frame X-ray luminosity in the 0.1-2.4
      keV band), $\mu\equiv\ln (M_{200c}$) (total mass measured within
      the overdensity of 200 with respect to the critical density). In
      addition to the redshift-dependent richness cut, we also
      discarded all clusters with richness below 25. The lowest and
      highest redshifts in the cleaned catalog are 0.047 and 0.682,
      respectively, but we restrict our analysis to the range $0.1 < z
      < 0.5$. After the richness cut, 1892 clusters lie within this
      redshift range.
     
     We applied the BOSS stellar mask to remove the areas affecting optical cluster detection. We considered that the optical completeness of the CODEX catalog above the applied richness cut does not to vary over the BOSS area  and modeled it using
      \begin{equation}
          \lambda_{50\%}(z) = \ln (17.2 + e^{\left( \frac{z}{0.32} \right)^{2}}),
      \end{equation}
which was obtained using the tabulations of \cite{redmapper}. We used an error function with the mean of $\lambda_{50\%}(z)$ and a $\sigma=0.2$, which reproduces the 75\% and 90\% quantiles of the distribution tabulated in  \cite{redmapper}. We used the probability of the optical detection of the cluster in SDSS data as
      \begin{equation}
          \label{eq.psdss}
          P^\mathrm{SDSS}(I|\lambda,z)=1-0.5\mathrm{erfc}\left(\frac{\lambda-\lambda_{50\%}}{0.2\sqrt{2}}\right).
      \end{equation}
     We verified the lack of sensitivity toward variations in the photometric depths using an agreement in the clustering signal between Northern and Southern Galactic Cap areas and using the tests for the presence of artificial correlation due to the bright stars.

    The RASS survey has large spatial inhomogeneities in its coverage; the limiting flux changes by a factor of 10. To properly account for the variation in the spatial distribution caused by it, we generated a random catalog in which the number of objects was six times higher than in the real catalog. 
    We divided the survey area into 100 zones of equal sensitivity $S$, with a 12\% step in flux sensitivity between the subsequent zones. We denote the sky area of these zones as $\Delta\Omega_{S}$. We computed the probability of cluster detection
   \begin{multline}
      \label{eqn:matrix}
         P(I|S,\mu,z,\nu) 
         = 
         \iiint \de l\true \de r_c \de \eta\ob P(I | \eta\ob , \beta(\mu), r_c )\\
         P(\eta\ob | \eta\true(l\true, S, z))  P(r_c, l\true | \mu, \nu, z),
    \end{multline}
    where $\eta$ denotes X-ray count (superscript "true" stands for predicted and "ob" for observed), $\nu\equiv\frac{\lambda - \langle\lambda | \mu,z\rangle}{\sigma_{\lambda | \mu}}$, and all the probabilities are described in detail in \cite{codex}. This modeling accounts for the effect of the X-ray shape ($r_c$, $\beta$) of the clusters and for the RASS sensitivity on the cluster selection, and predicts changes in the distribution of X-ray shapes based on the measured covariance of cluster properties \citep{cavaliere76, locuss, farahi, kaefer}. In this way, we accounted for the effect of the anticorrelation between the scatter of the X-ray luminosity and the optical richness and of the anticorrelation of the core radii of galaxy clusters and their scatter in luminosity. Our simulations reproduce the correct mix of shapes, luminosities, and richnesses of galaxy clusters. They were then used for the cluster detection modeling in all areas of the survey. 

\begin{figure}
   \centering
   \includegraphics[width=\hsize]{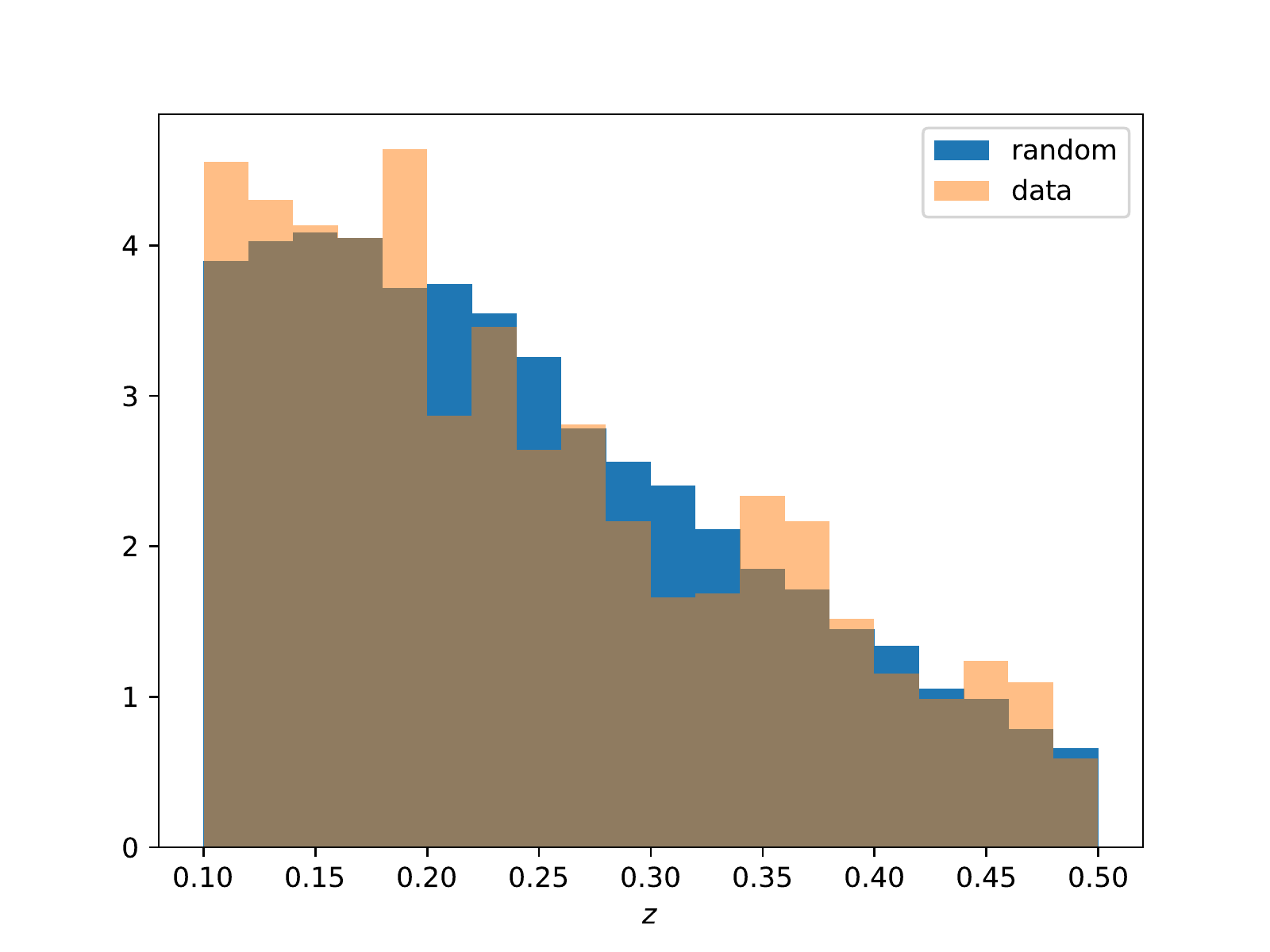}
      \caption{Redshift distribution ($dn/dz$) of the subset of the CODEX catalog used to compute the two-point correlation function along with the corresponding random catalog. The area of the two histograms is normalized to 1.}
         \label{fig:redshift_distr}
   \end{figure}

    We estimated the expected number of clusters in each bin of redshift as
    \begin{equation}
        \label{eqn:NC}
        \langle N(\Delta z) \rangle = \Delta\Omega_{S} \int_{\Delta z} \de z  \frac{\de V}{\de z \de \Omega}  (z) 
        \iint \de \mu \de \lambda \frac{\de n(\mu,\lambda, S, z)}{\de \mu \de \lambda \de V},
     \end{equation}
     where 
      \begin{multline}
      \label{eqn:hmf}
          \frac{\de n(\mu,\lambda, S, z)}{\de \mu \de \lambda \de V}
           =   P^\mathrm{RASS}(I | \lambda, z)P^\mathrm{SDSS}(I | \lambda, z) \\
           P(I|S,\mu,z,\nu) P(\lambda| \nu, \mu ) \frac{\de n(\mu, z) }{\de V \de\mu} \,.
      \end{multline}
      \fig{\ref{fig:redshift_distr}} shows a correspondence of the slope of the dn/dz distribution of the real clusters and the random catalog. They agree well.  
      \begin{figure}
         \centering
          \includegraphics[width=\hsize]{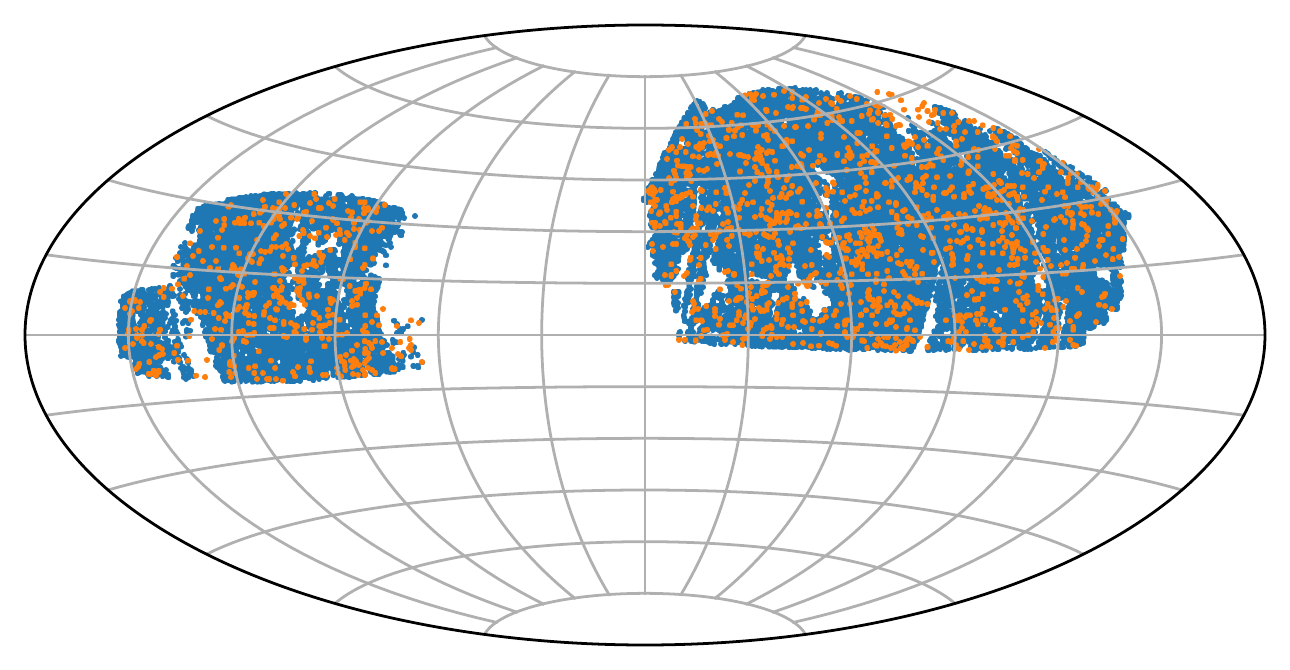}
          \caption{Sky footprint of the subset of the CODEX catalog used to compute the two-point correlation function. Orange points show the clusters, and blue points show the random points.}
         \label{fig:sky_footprint1}
   \end{figure}
   
\section{Analysis methods}
\label{sec:methods}

Our clustering analysis consists of first estimating the clustering of
the clusters by computing their two-point correlation function; next,
computing the expected bias of the sample from the cluster masses and
redshifts; then, computing a prediction for the clustering signal by
scaling a prediction for dark matter two-point correlation by the
square of the predicted bias; and finally, comparison between the
measured and predicted signals.

    We used the widely adopted Landy-Szalay estimator \citep{landyszalay} to estimate the two-point correlation function from our cluster sample. The estimator was constructed from pairs within the cluster catalog ($DD$), pairs within the random catalog ($RR$), and pairs between the two ($DR$):
   \begin{equation}
        \xi(\vec{r}) = \frac{n_R(n_R-1)}{n_D(n_D-1)}\frac{DD(\vec{r})}{RR(\vec{r})} - \frac{n_R(n_R-1)}{n_D n_R}\frac{DR(\vec{r})}{RR(\vec{r})} + 1,
        \label{eq:lz}
   \end{equation}
   where $\vec{r}$ is a vector encoding arbitrary separation bins, $n_D$ is the total number of clusters, and $n_R$ is the total number of random points. The simplest way to bin number counts is by their three-dimensional distance. This estimate is distorted by changes in cluster positions along the line-of- sight direction due to redshift caused by their peculiar velocities, however. These redshift space distortions can be minimized by estimating the so-called projected two-point correlation function \citep{davis83}. The projected two-point correlation function is the line-of-sight integral of two-point correlation function that has been binned in distance of the points along directions parallel and perpendicular to the line-of-sight direction:
   \begin{equation}
      w_p(r_p) = 2\int_{\pimin}^{\pimax}\xi(r_p, \pi)\de\pi.
      \label{eq:wp}
   \end{equation}
   The upper limit of the integral is selected in a way that increasing it would only increase noise and not the signal.
   
   We estimated the covariance matrix of the two-point correlation function using the jackknife method: we split the sample into $M$ subsections of the sky and computed a set of two-point correlation functions by excluding one subsection at a time. The covariance matrix element $C_{ij}$ was then computed as
   \begin{equation}
       C_{ij} = \frac{M-1}{M}\sum_{k=1}^M [ \xi_k(r_i) - \langle \xi_k(r_i)\rangle ] [ \xi_k(r_j) - \langle \xi_k(r_j)\rangle ],
   \end{equation}
   where $\langle\xi_k(r_j)\rangle$ is the mean over $M$ subsections. All the error bars we show for the two-point correlation function estimates are then the square root of the diagonal of this matrix, $\sigma_i = \sqrt{C_{ii}}$
   
   As discussed in the introduction, galaxy clusters are biased tracers of the total matter distribution, and their clustering bias is connected to dark matter masses of their host halos. The goal of our analysis is to compare mass-based cluster bias predictions to the actual bias measured from the two-point correlation function estimate. Thus a key ingredient in our analysis is the relation between cluster masses and their biases. It is possible to predict the clustering bias based on the mass function $n(M)$, that is, the number density of halos of a given mass, using the so-called peak background split (see, e.g., \citealp{mo_white96} and \citealp{sheth_tormen99}). In this approach it is  customary to consider masses in terms of the variance of the linear matter power spectrum $P(k, z)$,
   \begin{equation}
       \sigma^2(M,z) = 4\pi^2 \int_0^{\infty} P(K, z)W^2(k, M)k^2\de k,
   \end{equation}
   where $W(k, M)$ is the Fourier transform of a top-hat window function at $R=(3M/4\pi\overline{\rho}_M)^{1/3}$ that encloses mass $M$. Here $\overline{\rho}_M$ is the mean matter density in the Universe. For cleaner notation, we write in in \eqs{\ref{eq:nm}-\ref{eq:bias_tinker}} $\sigma$ instead of $\sigma(M,z)$. Now the halo mass function can be expressed as
   \begin{equation}
   \label{eq:nm}
   n(M)\de M = f(\sigma) \frac{\overline{\rho}_M}{M^2} \frac{\de\ln\sigma^{-1}}{\de\ln M}\de M,
   \end{equation}
   where $f(\sigma)$ is so-called multiplicity function. It is simply the fraction of mass contained in halos in a unit range of $\de\ln\sigma$. Using the ellipsoidal collapse model, \citet{sheth01} arrived at a parameterized multiplicity function,
   \begin{equation}
   \label{eq:multi_sheth_tormen}
   \begin{split}
       f_{\mathrm{ST}}(\sigma; A, a, p) &= A \sqrt{\frac{2}{\pi}}\left[1 + \left(\frac{\sigma^2}{a^2\delta_c^2}\right)^{p}\right]\cdots \\
       \cdots &\left(\frac{\sqrt{a}\delta_c}{\sigma}\right)\exp\left[-\frac{a}{2}\frac{\delta_c^2}{\sigma^2} \right],
    \end{split}
    \end{equation}
   where$\delta_c = 1.686$ is the critical density for halo collapse. To improve the compatibility with $N$-body simulations, \citet{bhattacharya} introduced an additional ad hoc parameter $q$, which is a function of redshift, and changed
   \begin{equation*}
   \left(\frac{\sqrt{a}\delta_c}{\sigma}\right) \rightarrow \left(\frac{\sqrt{a}\delta_c}{\sigma}\right)^{q}
    \end{equation*}
   and consequently allowed parameters $(A, a, p)$ in \eq{\ref{eq:multi_sheth_tormen}} to evolve with redshift. By following the procedure of computing ratio of conditional and unconditional mass function (from \citealp{sheth_tormen99}), we can express the halo bias using the parameters $(a, p, q)$ of the halo mass function model,
   \begin{equation}
   \label{eq:bias_comparat}
        \begin{split}
           b(\sigma, a, p, q) &= 1 + \frac{a(\delta_c/\sigma)^2 - q}{\delta_c} \\
           &\cdots + \frac{2p/\delta_c}{1 + (a(\delta_c/\sigma)^2)^p}
       \end{split}
   .\end{equation}
   The parameters of this model can be calibrated by identifying dark matter halos in $N$-body simulations and estimating the clustering amplitudes of halo populations in different mass bins and comparing them to the total matter distribution. Instead of semianalytical formulas based on halo mass function, we can also consider purely empirical fitting functions, as was done by \citet{tinker10}, for instance. They allow a more flexible functional form,
   \begin{equation}
   \label{eq:bias_tinker}
      b(\sigma, A, a, B, b, C, c) = 1 - A\frac{(\delta_c/\sigma)^a}{(\delta_c/\sigma)^a + \delta_c^a} + B(\delta_c/\sigma)^b + C(\delta_c/\sigma)^c
   .\end{equation}
   Again, the parameters $(A, a, B, b, C, c)$ were fixed by fitting $N$-body simulations. In section \ref{sec:validation} we consider both mass-function-based and empirical bias-to-mass models.
   
    With equations \ref{eq:bias_comparat} and \ref{eq:bias_tinker}, we can predict the clustering bias of a halo population of a given mass at a given redshift. The simplest way to use this to estimate the clustering bias within a cluster sample is to compute the mean bias within the sample as
    \begin{equation}
    \label{eq:mean_bias}
        \overline{b} = \frac{1}{n_D}\sum_{i=1}^{n_d} b(M_i, z_i)g(z_i),
    \end{equation}
    where $n_D$ is again the number of clusters in the sample, and $b(M_i, z_i)$ is the predicted bias for cluster with mass $M_i$ at redshift $z_i$. Factor $g(z_i)$ is a redshift-dependent correction for the growth of structure, which is required because we include estimates across a wide redshift range. The correction factor is defined as
   \begin{equation}
        g(z) = \sqrt{\frac{\wpdm(z, r)}{\wpdm(0, r)}}.
       \label{eq:gz}
   \end{equation}
   Here $\wpdm(z,r)$ is the dark matter projected two-point correlation function at a redshift $z$ and a scale $r$. Thus in principle, factor $g(z)$ is also a function of scale $r,$ but following \citet{allevato11}, we approximated it simply by $D_1(z)/D_1(0)$, where $D_1(z)$ is the growth function \citep[see, e.g.,][]{eisensteinhu1}. If the mass estimates and mass-to-bias conversion are correct, we expect the bias-scaled dark matter two-point correlation function at $z=0$, that is, $\overline{b}^{\; 2}\xidm(0, r)$, to agree with the measured two-point correlation function. Throughout this paper all the predictions for the dark matter two-point correlation functions are Fourier transforms from linear dark matter power spectra computed based on \citet{eisensteinhu2} and \citet{eisensteinhu1} (referred to as Eisenstein \& Hu hereinafter)  at redshift $z=0$. We also tried using power spectra from the Code for Anisotropies in the Microwave Background (CAMB) \citep{camb}, but the differences were negligible and our pipeline implementation is faster by roughly a factor of ten when using Eisenstein \& Hu (which is important for cosmological parameter space scanning). Thus all the results we present are based on Eisenstein \& Hu.
   
   We used the CosmoBolognaLib\footnote{\url+https://github.com/federicomarulli/CosmoBolognaLib+} C++/python library \citep{cosmobol} to compute the pair counts needed to estimate the two-point correlation function and also to divide sky into subregions to compute the covariance matrix from the corresponding jackknife sample. Dark matter two-point correlation function predictions, mass-to-bias conversion, and all the other theoretical cosmology-dependent computations were performed using the COLOSSUS\footnote{\url+https://bitbucket.org/bdiemer/colossus+} python library \citep{colossus}.

\section{Comparison with simulations}
\label{sec:validation}

    To validate our analysis pipeline, we ran it using a dark matter
    halo catalog from the Huge MultiDark Planck (HMDPL) simulation
    \citep{mdpl2, klypin16}. This is a dark matter only $N$-body
    simulation in a $4h^{-1}\mathrm{Gpc}$ box with a cosmology that is
    consistent with the results of \cite{planck2013}. In the case of
    the simulated catalog, we know the underlying cosmology and halo
    masses exactly, therefore we expect to be able to recover the halo
    bias at a high accuracy. The catalog we used in our test is a
    light cone obtained from the HMDPL simulation. This light cone
    covers the full sky in the redshift range $0.00 < z < 1.8$. In our
    clustering analysis we masked galactic latitudes $g_\mathrm{lat} <
    10$ and picked a subset of halos with redshifts $0.1 < z < 0.5$
    and masses $M_{200c} > 4.8 \times 10^{14}h^{-1}\Msun$, which
    produces a sample with similar masses and redshift as in our CODEX
    sample. After these selections, the catalog contained 6350
    halos. We estimated the covariance matrix for the two-point
    correlation function by splitting the sky area into 35 jackknife
    subsections.
   
\begin{figure}
    \centering
    \includegraphics[width=\hsize]{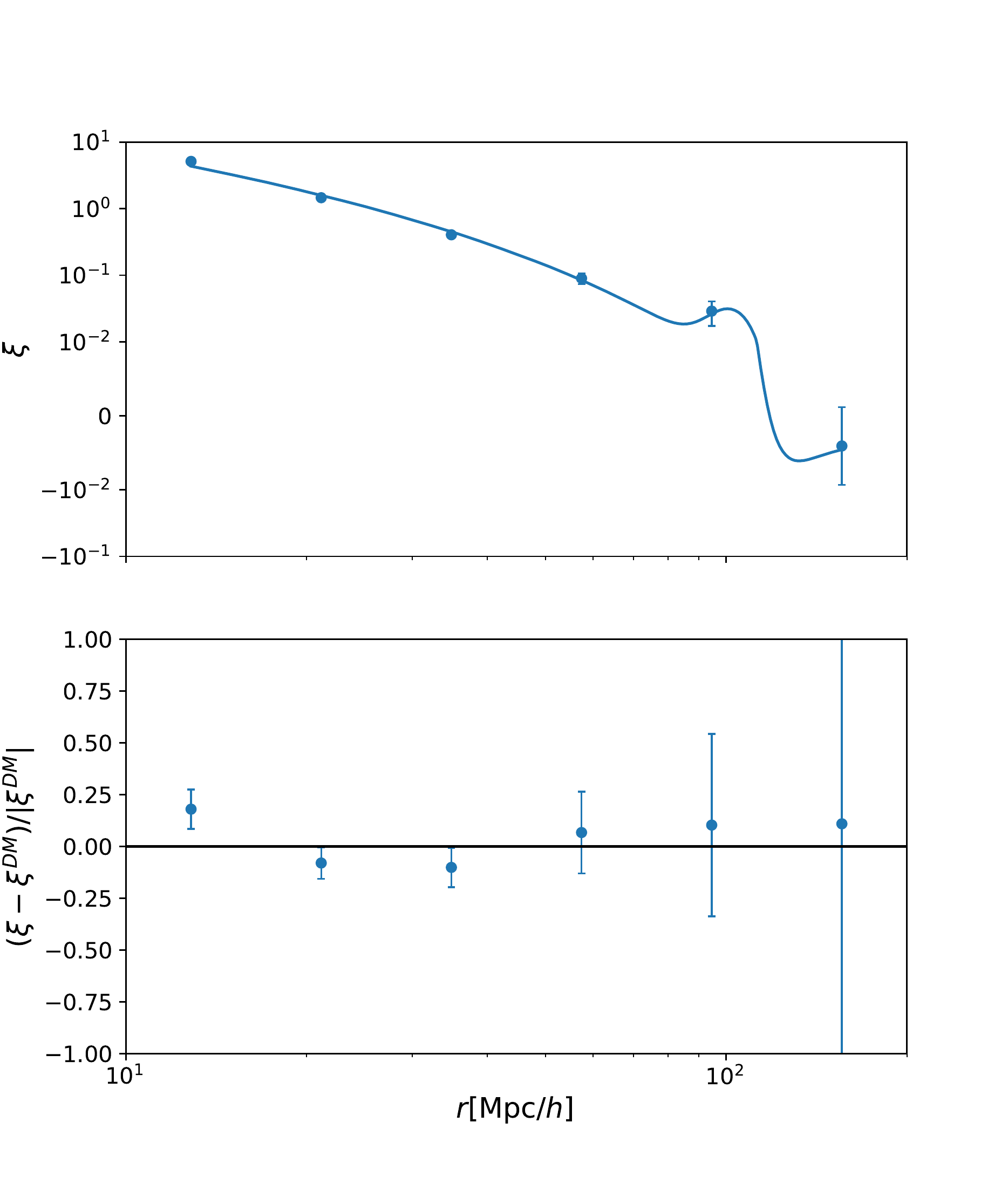}
    \caption{One-dimensional two-point correlation function from HMDPL halos. Top panel: Data points are the two-point correlation function estimate from the halo distribution. The solid curve is the predicted dark matter two-point correlation function scaled by $\overline{b}^{\; 2}$. Bottom panel: Relative difference between the measured and predicted two-point correlation functions. }
    \label{fig:mdpl2_1d}
   \end{figure}

   The HMDPL catalog contains halo redshifts with and without their
   peculiar motion. To validate our method in a scenario that is free of
   complications from redshift-space effects and binning in two
   dimensions we first computed the one-dimensional two-point
   correlation function $\xi(r)$ using purely cosmological
   redshifts. In this case, the pair counts were binned simply in the
   three-dimensional distance of the points. We computed the two-point
   correlation function estimate in six bins spaced logarithmically
   over scales of $10 h^{-1}\Mpc < r < 200 h^{-1}\Mpc$. Figure
   ~\ref{fig:mdpl2_1d} shows a comparison between the measured halo
   two-point correlation function and a dark matter two-point
   correlation prediction scaled by the square of the halo bias. The
   bias estimate was obtained from the known halo masses using
   \eq{\ref{eq:mean_bias}} and the $b(M)$ calibration from
   \citet{comparat17}, which is based on the HMDPL simulation. In
   addition to \citet{comparat17} we tested three other models for $b(M)$ from
   \citet{sheth01}, \citet{tinker10}, and \citet{bhattacharya}. The
   resulting biases are presented in
   \tb{\ref{tab:biascomparison}}. They can be compared to the bias
   estimate we obtained by directly fitting the bias factor, using its
   definition as
   \begin{equation}
       b = \sqrt{\frac{\xi(r)}{\xidm(r)}}.
   \end{equation}
   We used our jackknife covariance matrix estimate to obtain a least-squares fit for the bias factor, which in this case gives $b=4.33 \pm 0.07$. \citet{comparat17} predicted a value of $\overline{b} = 4.29,$ which agrees best with the predicted bias as expected because it was calibrated on the same simulation that we used for validation. For the rest of this paper we use the model of \citet{comparat17} as our baseline model and treat the deviations from this as the source of the systematic uncertainty. 
       \begin{table}
        \caption{Biases predicted using different models $b(M)$.}
        \label{tab:biascomparison}
        \centering
            \begin{tabular}{c c}
            \hline\hline
            $b(M)$ & $\overline{b}$ \\
            \hline
            \citet{sheth01} & 4.04 \\
            \citet{tinker10} & 4.79 \\
            \citet{bhattacharya} & 3.68 \\
            \citet{comparat17} & 4.29\\
            \hline
        \end{tabular}
    \end{table}
   
\begin{figure}
    \centering
    \includegraphics[width=\hsize]{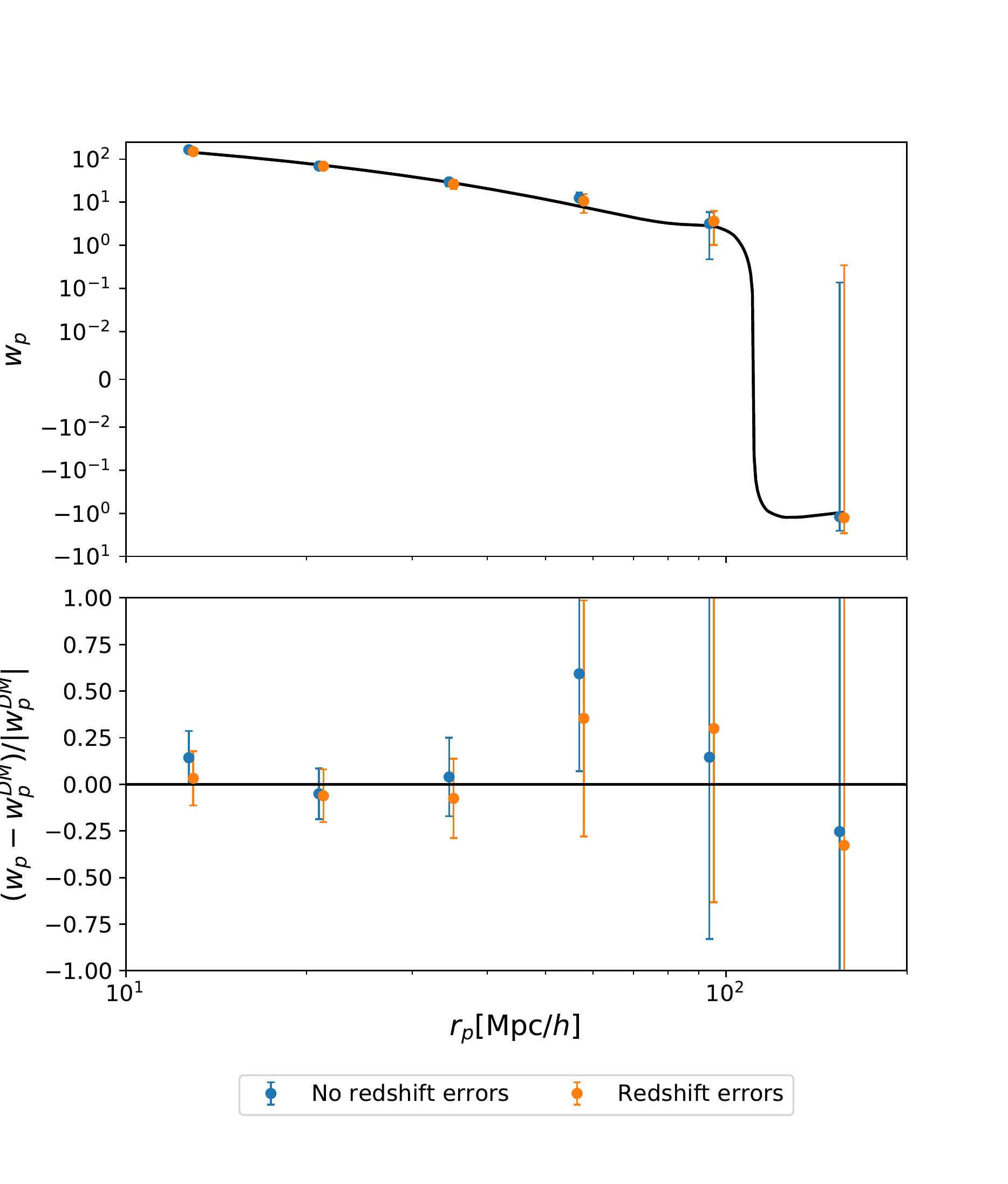}
    \caption{Effect of redshift errors on the projected two-point correlation function estimate. Top panel: Projected two-point correlation function from HMDPL halos. Blue data points are the two-point correlation function estimate from the halo distribution without redshift errors, and orange data points are this estimate including redshift errors. The solid curve shows the predicted dark matter two-point correlation function scaled by $\overline{b}^{\; 2}$ computed from the halo masses. Bottom panel: Relative difference between the measured and predicted two-point correlation functions.}
    \label{fig:wp_mdpl2_phz_err}
   \end{figure}

  We then studied the effect of including halo peculiar velocities in
  their redshifts. In this case, the appropriate clustering statistics
  is the projected two-point correlation function introduced in
  Sec.~\ref{sec:methods}. Our dark matter prediction is purely
  isotropic and obtained from the one-dimensional two-point
  correlation function by setting $\xidm (r_p, \pi) =
  \xidm\left(\sqrt{r_p^2 + \pi^2}\right)$. The binning in $r_p$
  direction was the same as in distance $r$ for our one-dimensional
  $\xi(r)$ estimate. We chose $\pimin = 0 h^{-1}\Mpc$ and $\pimax =
  120 h^{-1}\Mpc$ for the integral in \eq{\ref{eq:wp}}. The result of
  this integral is the projected two-point correlation function
  $w_p(r_p)$ shown in \fig{ \ref{fig:wp_mdpl2_phz_err}} along with the
  corresponding bias-scaled dark matter prediction. Because the halo
  sample is still the same, the mass-based bias prediction does not
  change compared to the one-dimensional case, but fitting the bias
  gives a slightly higher value of $b = 4.41 \pm 0.11$. The difference
  of the predicted and measured bias is slightly larger than the
  $1\sigma$ uncertainty in the latter. The difference is $3 \%$ of the
  predicted bias, which is well within the estimated uncertainty of
  the $b(M)$ model \citep{comparat17}.
   
   In real data, the radial cluster distances are distorted by
   redshift measurement errors in addition to their peculiar
   velocities. For a discussion of this effect, see \citet{estrada09},
   for instance. To model the effect, we measured the redshift error
   distribution from the CODEX clusters. Spectroscopic redshifts are
   available for a subset of CODEX clusters from the SDSS IV DR16
   release of the SPIDERS cluster catalog \citep{ahumada19}.
   Descriptions of the survey area and spectroscopic redshift
   assignment are provided in \citet{clerc20} and \citet{Kirkpatrick},
   respectively.  For areas outside of the SPIDERS footprint,
   redshifts were collected from public SDSS III data and several
   Nordic Optical Telescope (NOT) programs (PI A. Finoguenov, NOT
   Program IDs: 48-025, 52-026, 53-020, 51-034).  The redshift
   assignment follows the same procedure as we used for the DR16
   catalog. The number of clusters with spectroscopic redshifts is
   1223, or 65\% of the entire sample. We estimated the redshift error
   distribution as the difference between spectroscopic and
   photometric redshift for each cluster. We then fit this
   distribution with a Gaussian function of width $\sigma_z$ and used
   this function to draw a random redshift error for each halo in the
   simulated catalog. We obtained $\sigma_z = 0.0071$ in the range
   $0.1 < z <0.5$. The effect this has on measured projected two-point
   correlation function is shown in
   \fig{\ref{fig:wp_mdpl2_phz_err}}. The corresponding measured bias
   of $4.29 \pm 0.07$, which is compatible with the bias without
   redshift errors and also with the predicted bias within the
   statistical errors. Thus we do not expect the measurement of the
   projected two-point correlation function to be biased by the
   redshift measurement errors in the CODEX catalog.
 
\section{Clustering of CODEX clusters}
    \label{sec:codex_clustering}
\begin{figure}
    \centering
    \includegraphics[width=\hsize]{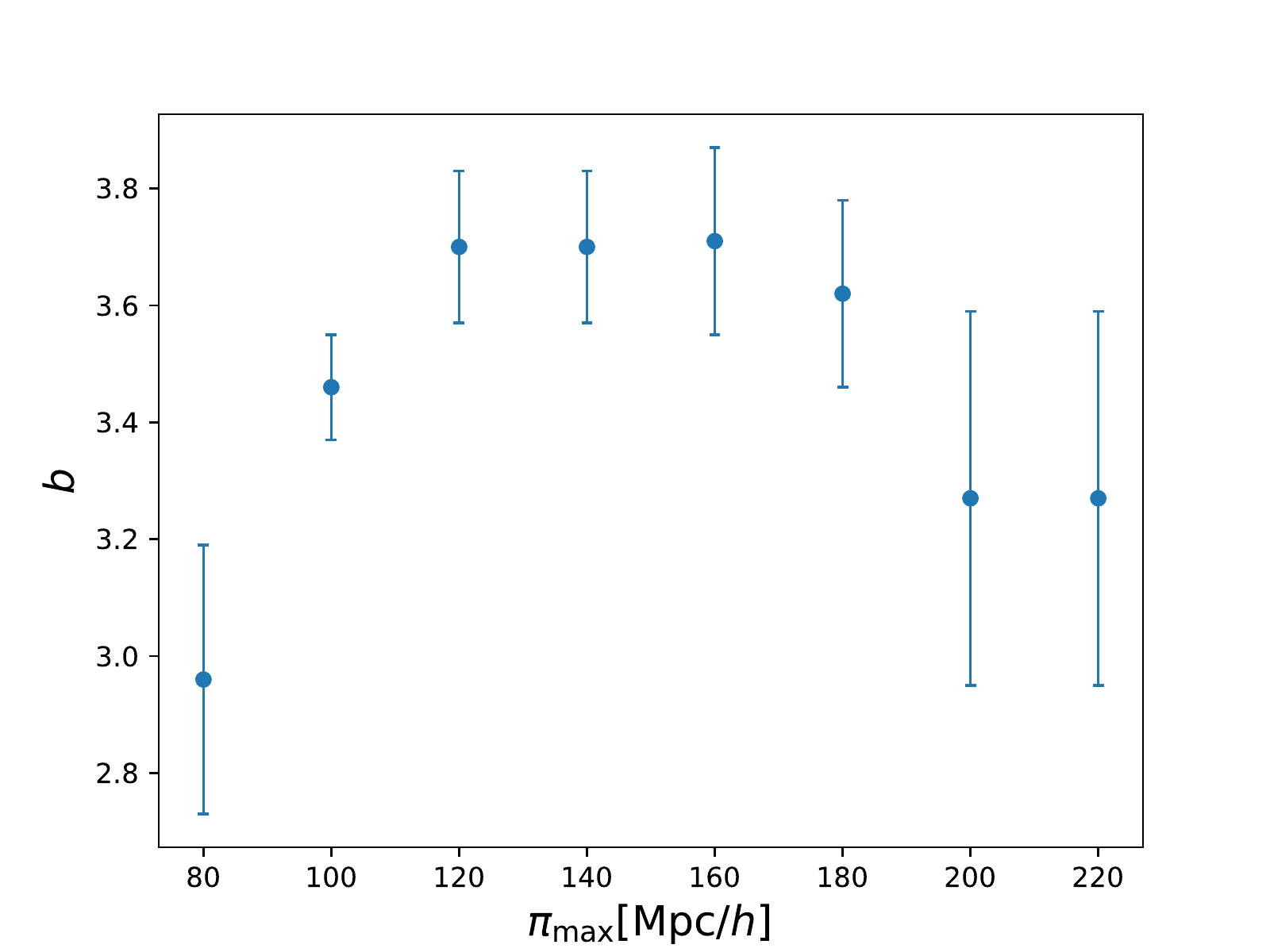}
    \caption{Effect of $\pimax$ to the clustering amplitude quantified via fitted bias $b$.}
    \label{fig:bias_pimax}
    \end{figure}

    Because the predicted halo bias is compatible with the measured bias for the simulated halos, we proceeded to apply the same analysis to the CODEX cluster catalog. We again estimated the projected two-point correlation function. We chose five logarithmically spaced bins over scales of $10h^{-1}\Mpc < r_p < 200h^{-1}\Mpc$ and integrated over scales of $0h^{-1}\Mpc < \pi < 120h^{-1}\Mpc$. We also tried using other values for $\pimax$ , but $120h^{-1}\Mpc$ maximizes the clustering amplitude with minimum noise. Figure~\ref{fig:bias_pimax} shows the fitted bias as a function of $\pimax$. To derive the covariance matrix estimates, we split the sky area into 16 jackknife subsections. 
    
\begin{figure}
    \centering
    \includegraphics[width=\hsize]{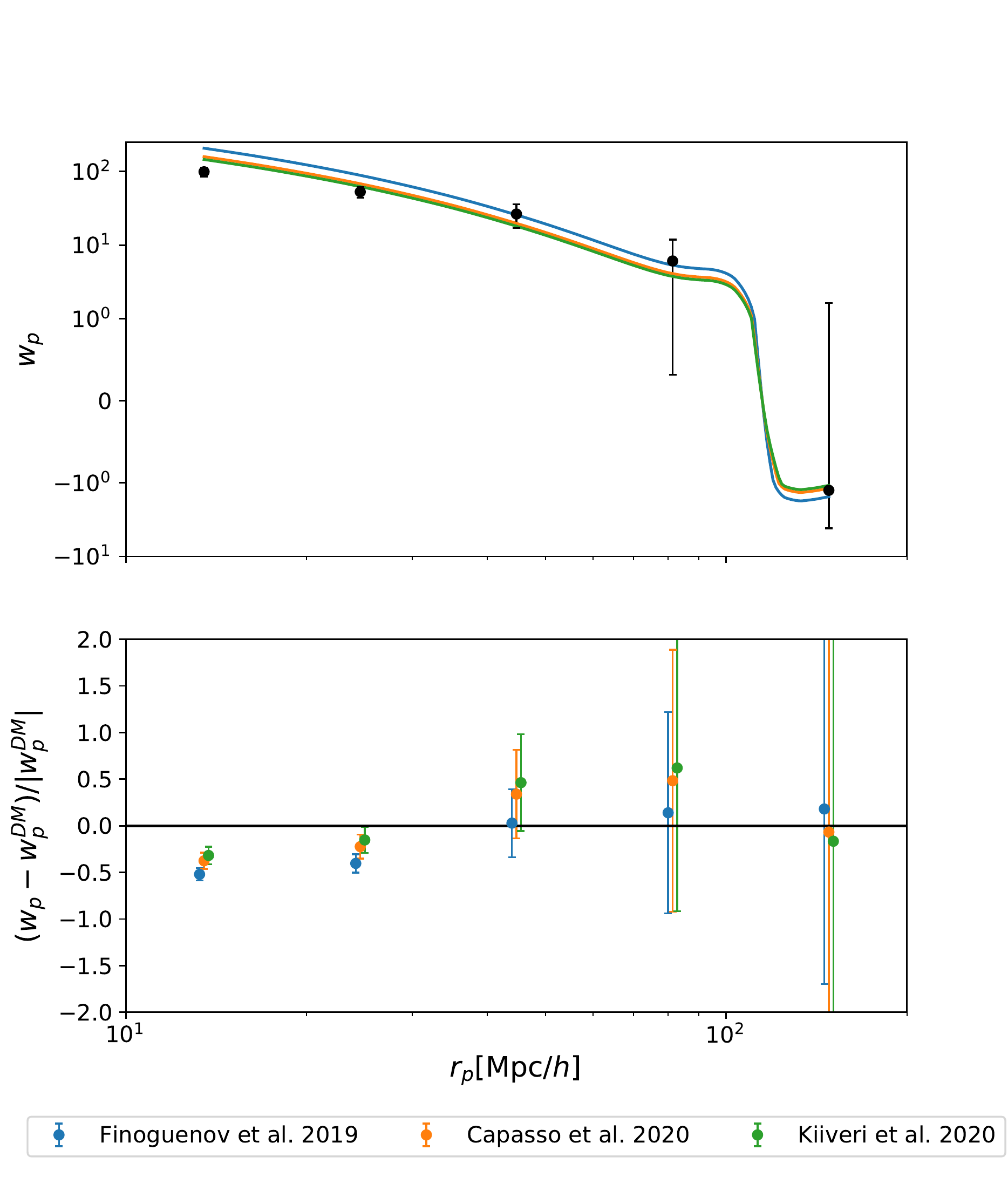}
    \caption{Comparison of different cluster mass estimates. Top panel: projected two-point correlation function in redshift range $0.1 < z < 0.5$. Data points are two-point correlation function estimate from CODEX clusters. The solid curves show the predicted dark matter two-point correlation function scaled by $\overline{b}^{\; 2}$ computed using the different halo mass estimates. Bottom panel: Relative difference between the measured and predicted two-point correlation functions. The data points have been shifted horizontally for clarity.}
    \label{fig:codex_masscomparison}
    \end{figure}

    \fig{\ref{fig:codex_masscomparison}} shows the projected two-point
    correlation function obtained from the CODEX catalog compared with
    dark matter two-point correlation functions scaled with the
    predicted bias factors based on three different mass
    estimates. Two estimates come from cluster X-ray luminosities
    ($L_X$) using two different scaling relations, one adopted in the
    CODEX main paper \citep{codex}, and the other from
    \cite{capasso20}. The third estimate comes from the cluster
    richness, using the summary of weak-lensing richness calibrations
    presented in \cite{richness-mass-kimmo}. In
    \tb{\ref{tab:bias_masscomparison}} we show the mass-based bias
    predictions. These can be compared to the bias estimate obtained
    by direct fitting, which in this case is $b = 3.70 \pm 0.13$. The
    richness-based mass estimates predict a bias of $\overline{b} =
    4.33,$ which is closest to the measured value, the difference
    being $17\%$ of the fitted value. The superiority of the
    richness-based masses is most likely due to the better quality of
    its measurement compared to $L_X$ based on a few counts in RASS
    data. Thus, we selected richness to be our baseline case in the
    following sections and used the deviating results from other mass
    estimates to illustrate an associated systematic uncertainty. 
    \begin{table}
    \caption{Predicted biases using different mass estimates.}
    \label{tab:bias_masscomparison}
    \centering
    \begin{tabular}{c c c}
    \hline\hline
    Mass proxy & Calibration & $\overline{b}$ \\
    \hline
    $M_{200c} (L_x)$ & \citet{codex} & 5.16 \\
    $M_{200c} (L_x)$ & \citet{capasso20} & 4.52 \\
    $M_{200c}$($\lambda$) & \citet{richness-mass-kimmo} & 4.33 \\
    \hline
    \end{tabular}
    \end{table}
    To verify the robustness of our results against the $r_p$ range, we
    also ran the analysis in the ranges $10h^{-1}\Mpc < r_p <
    100h^{-1}\Mpc$ and $20h^{-1}\Mpc < r_p < 200h^{-1}\Mpc$. The
    results are listed in \tb{\ref{tab:rp_comparison}}. Increasing the
    lower limit has a stronger effect on the fitted bias than
    decreasing the upper limit, but all the results are compatible
    within the statistical uncertainty.
        \begin{table}
        \caption{Fitted bias using different $r_p$ ranges.}
        \label{tab:rp_comparison}
        \centering
            \begin{tabular}{c c c}
            \hline\hline
            Range & $b$ \\
            \hline
            $10h^{-1}\Mpc < r_p < 200h^{-1}\Mpc$ & $3.70 \pm 0.13$\\
            $10h^{-1}\Mpc < r_p < 100h^{-1}\Mpc$ & $3.68 \pm 0.33$\\
            $20h^{-1}\Mpc < r_p < 200h^{-1}\Mpc$ & $3.99 \pm 0.25$\\
            \hline
        \end{tabular}
    \end{table}
    
\begin{figure}
   \centering
   \includegraphics[width=\hsize]{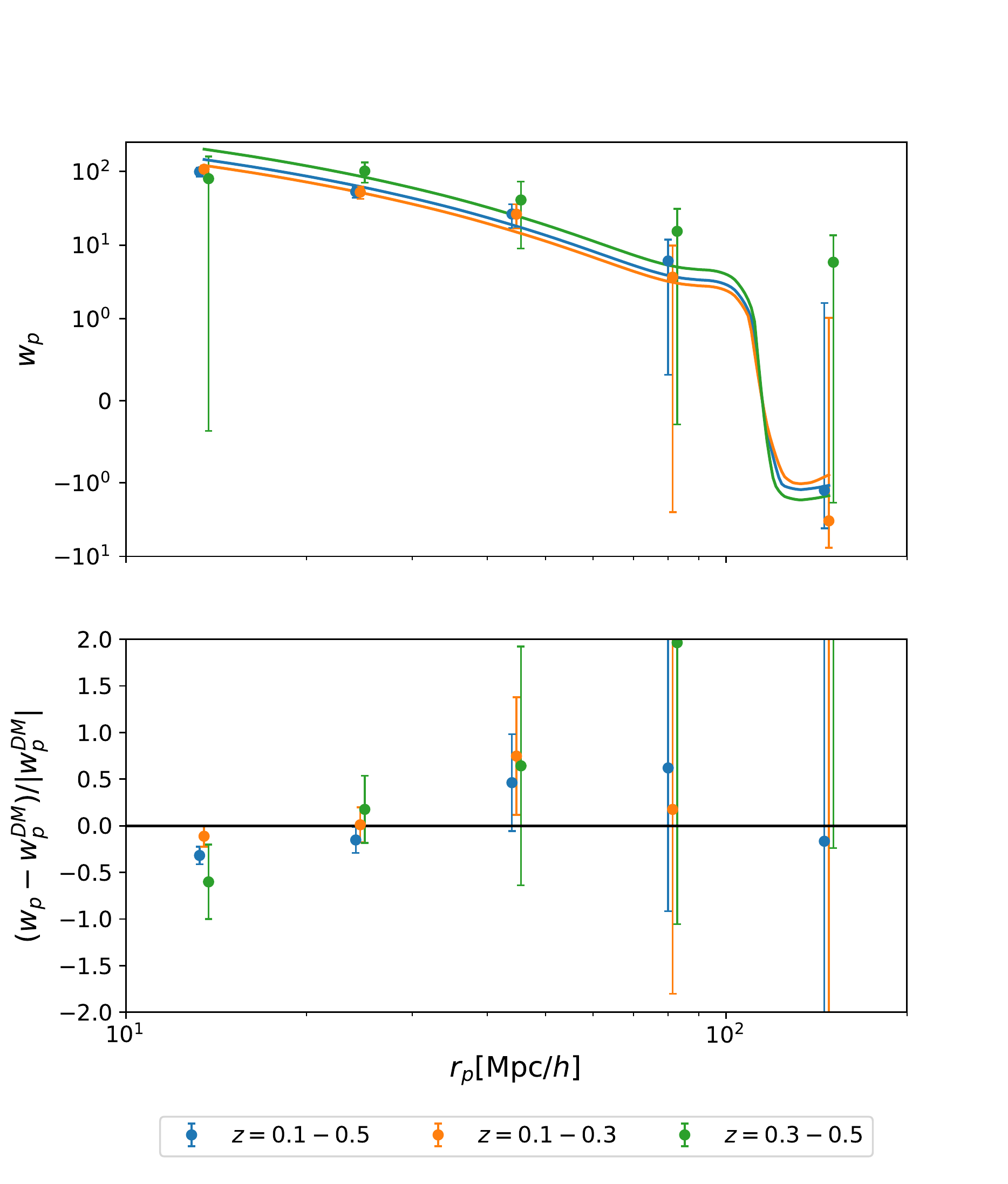}
      \caption{Comparison of different redshift ranges. Top panel: Projected two-point correlation function in three redshift ranges $0.1 < z < 0.5$, $0.1 < z < 0.3$ and $0.3 < z < 0.5$. Data points are the two-point correlation function estimates from the CODEX clusters. The solid curves in corresponding colors show the predicted dark matter two-point correlation function scaled by $\overline{b}^{\; 2}$ computed within each cluster subsample. Bottom panel: Relative difference between the two-point correlation function estimate and the corresponding prediction. The data points have been shifted horizontally for clarity.}
         \label{fig:codex_zcomparison}
   \end{figure}

    We also studied the effect of redshift binning on
    clustering. Figure~\ref{fig:codex_zcomparison} compares the CODEX
    clusters and predicted bias-scaled dark matter two-point
    correlation functions in lower $0.1 < z <0.3$ and higher $0.3 < z
    < 0.5$ redshift samples, and for reference, also in the full range
    of $0.1 < z < 0.5$. The corresponding comparison between predicted
    and fitted biases is presented in
    \tb{\ref{tab:bias_zcomparison}}. The difference between the
    predicted and measured bias for the low- and high-redshift samples
    is 5\% and 8\% of the measured bias, respectively. The measured
    high-redshift two-point correlation function appears to deviate
    from the prediction at the largest scales. This might be a
    residual from the limitations caused by modeling the survey
    effects at high redshifts. We give a more detailed account of this
    matter in Appendix~\ref{sec:systematics}. The predicted and
    measured biases are also higher for the high-redshift sample. This
    is expected because with increasing redshift, only massive
    clusters can be detected in RASS. This effect is further
    reinforced by our richness cut, which discards increasingly rich
    clusters at higher redshifts.
        \begin{table}
        \caption{Comparison of the measured  and predicted biases in different redshift ranges. Column $b$ is the measured bias and $\overline{b}$ is the mass-based prediction. Column $n_D$ shows the number of clusters in each redshift bin.}
        \label{tab:bias_zcomparison}
        \centering
            \begin{tabular}{c c c c}
            \hline\hline
            Redshift range &  $n_D$ & $\overline{b}$ & $b$\\
            \hline
            $0.1 < z < 0.5$ & 1892 & 4.33 & $3.70 \pm 0.13$\\
            $0.1 < z < 0.3$ & 1250 & 3.95 & $3.78 \pm 0.10$\\
            $0.3 < z < 0.5$ & 642  & 5.08 & $4.71 \pm 0.66$\\
            \hline
        \end{tabular}
    \end{table}
   
   A common way to characterize the amplitude of the cluster two-point correlation function is to fit a power law,
   \begin{equation}
      \xi(r) = \left( \frac{r}{r_0}\right)^{-\gamma},
   \end{equation}
   which is found to be a good approximation at the scales $\lesssim 100 \Mpc/h$ (see, e.g., \citealp{peacock92}). The scale at which the correlation function crosses unity, $r_0$, is called the correlation length, and determining its value has been the goal of many galaxy cluster studies. \citet{collins00} measured the correlation length from another smaller X-ray selected sample provided by the ROSAT-ESO Flux-Limited X-ray \citep[REFLEX,][]{bohringer01} survey. The value they obtained is $r_0 = 18.8 \pm 0.9,$ with a slope of $\gamma = 1.83
  ^{+0.15}_{-0.08}$ over the range of $4 \Mpc/h < r < 40 \Mpc/h$. To compare with this result, we performed a least-squares fit for parameters $r_0$ and $\gamma$. The results for our three redshift ranges are listed in \tb{\ref{tab:power_law}}. We measure a slightly steeper slope, especially for the high-redshift sample, but within the errors all the results are compatible with those of \citeauthor{collins00} (2000).
  \begin{table}
    \caption{Fitted power-law parameters for different redshift ranges.}
    \label{tab:power_law}
    \centering
    \begin{tabular}{c c c}
        \hline\hline
        Redshift range & $r_0$ & $\gamma$ \\
            \hline
            $0.1 < z < 0.5$ & $18.7 \pm 1.1$ & $1.98 \pm 0.14$ \\
            $0.1 < z < 0.3$ & $18.2 \pm 1.1$ & $2.13 \pm 0.15$ \\
            $0.3 < z < 0.5$ & $18.1 \pm 1.3$ & $1.97 \pm 0.14$ \\
            \hline
    \end{tabular}
  \end{table}

\section{Cosmology}
\label{sec:cosmology}
   As a of a proof-of-concept study, we performed a Markov chain Monte
   Carlo (MCMC) sampling \citep[implemented using the emcee
     library\footnote{\url+https://github.com/dfm/emcee+},][]{emcee}
   of the matter density parameter $\Om$ and power spectrum amplitude
   $\sigma_8$ within the $\Lambda$CDM model by comparing the two-point
   correlation function obtained from cluster distribution (halo
   distribution in case of simulations) and one obtained by scaling
   dark matter two-point correlation function with predicted bias. All
   the other cosmological parameters were fixed at values based on
   Wilkinson Microwave Anisotropy Probe (WMAP) nine-year results
   \citep{wmap9} in the case of CODEX clusters (we later combine the
   posterior with one based on cluster mass function and WMAP
   nine-year cosmology) or Planck 2015 results in the case of HMDPL
   halos (the simulation was run using this cosmology). We restricted
   ourselves to spatially flat cosmologies so that the dark energy
   density parameter is determined by the matter density parameter:
   $\Ode = 1 - \Om$ for each value of $\Om$. We assumed a Gaussian
   likelihood,
   \begin{equation}
       \ln{\mathcal{L}} = -\frac{1}{2}\left[\left(\xivec - \overline{b}^{\; 2}\xidmvec \right)^T \mathbf{C}^{-1}\left(\xivec - \overline{b}^{\; 2}\xidmvec \right) + c\right],
   \end{equation}
   where $c$ is a constant. The predicted two-point correlation
   function $\xidmvec$ and the mean bias $\overline{b}$ are both
   functions of cosmological parameters and thus updated at each MCMC
   step. For simplicity, we ignored the effect of errors in cluster
   masses on $\overline{b}$ and on the likelihood.
   
\begin{figure}
   \centering
   \includegraphics[width=\hsize]{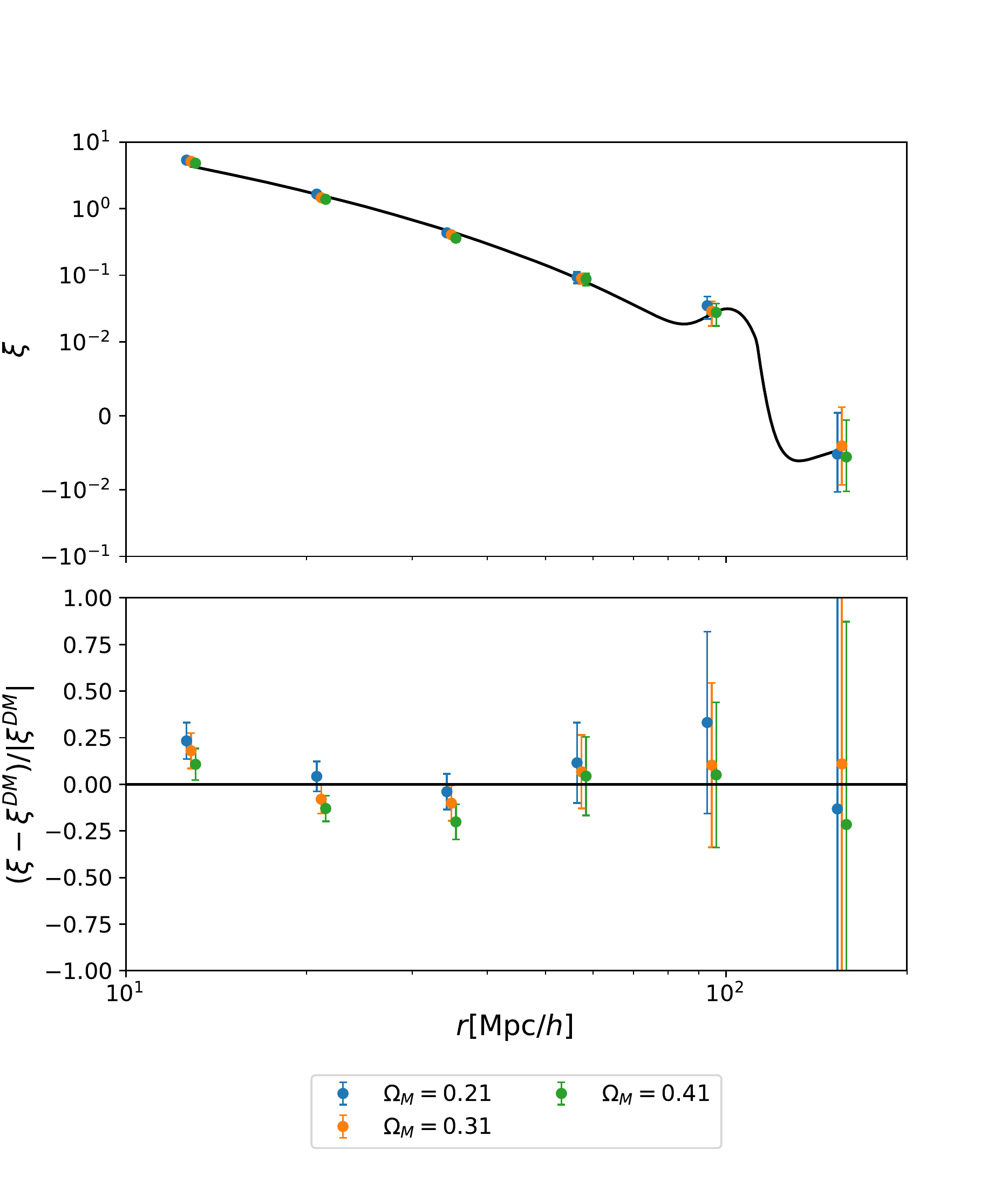}
      \caption{Effect of changing $\Om$ on the two-point correlation function estimate. Top panel: Bias-scaled dark matter prediction for HMDPL cosmology $\Om = 0.307115$ (solid curve) and two-point correlation function estimates obtained using this same value and values $\pm 0.1$. Bottom panel: Relative difference between the estimates and the dark matter prediction. The data points have been shifted horizontally for clarity.}
         \label{fig:omegam}
   \end{figure}

   The estimated two-point correlation function $\xivec$ changes with
   cosmology as well as with the cosmology-dependent transformation of
   redshifts to distances; in principle, it should therefore be
   reestimated at each cosmology. This effect is small,
   however. Figure~\ref{fig:omegam} compares two-point correlation
   function estimates from HMDPL halos using drastically differing
   values of $\Om$. The fitted biases are presented in
   \tb{\ref{tab:bias_omegamcomparison}}. Both of the extreme values
   are compatible with the central value within the statistical
   errors. To take the difference into account, we modeled the effect
   of comparing estimated and predicted two-point correlation
   functions computed at different cosmologies (so-called geometrical
   distortions, GD) following \citet{marulli12}. Distances
   perpendicular and parallel to the line of sight, $r_p$ and $\pi$,
   are related in the two different cosmologies, labeled 1 and 2, by
  \begin{equation}
      r_{p,1} = \frac{D_{A,1}(z)}{D_{A,2}(z)}r_{p,2}; \quad \pi_1 = \frac{H_2(z)}{H_1(z)}\pi_2,
  \end{equation}
   where $D_A(z)$ is the angular diameter distance and $H(z)$ is the Hubble parameter at redshift $z$. To compare theoretical predictions at varying  cosmologies with the measurement at the fiducial cosmology at scales $(r_p, \pi),$ we therefore evaluated them at scales $\left(D_{A}(z)/D_{A,f}(z)r_p, H(z)_f/H(z)\pi\right)$, where $f$ refers to the fiducial cosmology, and we took $z$ to be the mean redshift of the sample.
       \begin{table}
        \caption{Comparison of fitted biases using different values of $\Om$ while estimating the two-point correlation function. These should be compared to the mass-based prediction of $\overline{b} = 4.29$.}
        \label{tab:bias_omegamcomparison}
        \centering
            \begin{tabular}{c c}
            \hline\hline
            $\Om$ & $b$ \\
            \hline
            0.21 & $4.48 \pm 0.08$\\
            0.31 & $4.33 \pm 0.07$ \\
            0.41 & $4.18 \pm 0.09$ \\
            \hline
        \end{tabular}
    \end{table}
    
\begin{figure}
   \centering
   \includegraphics[width=\hsize]{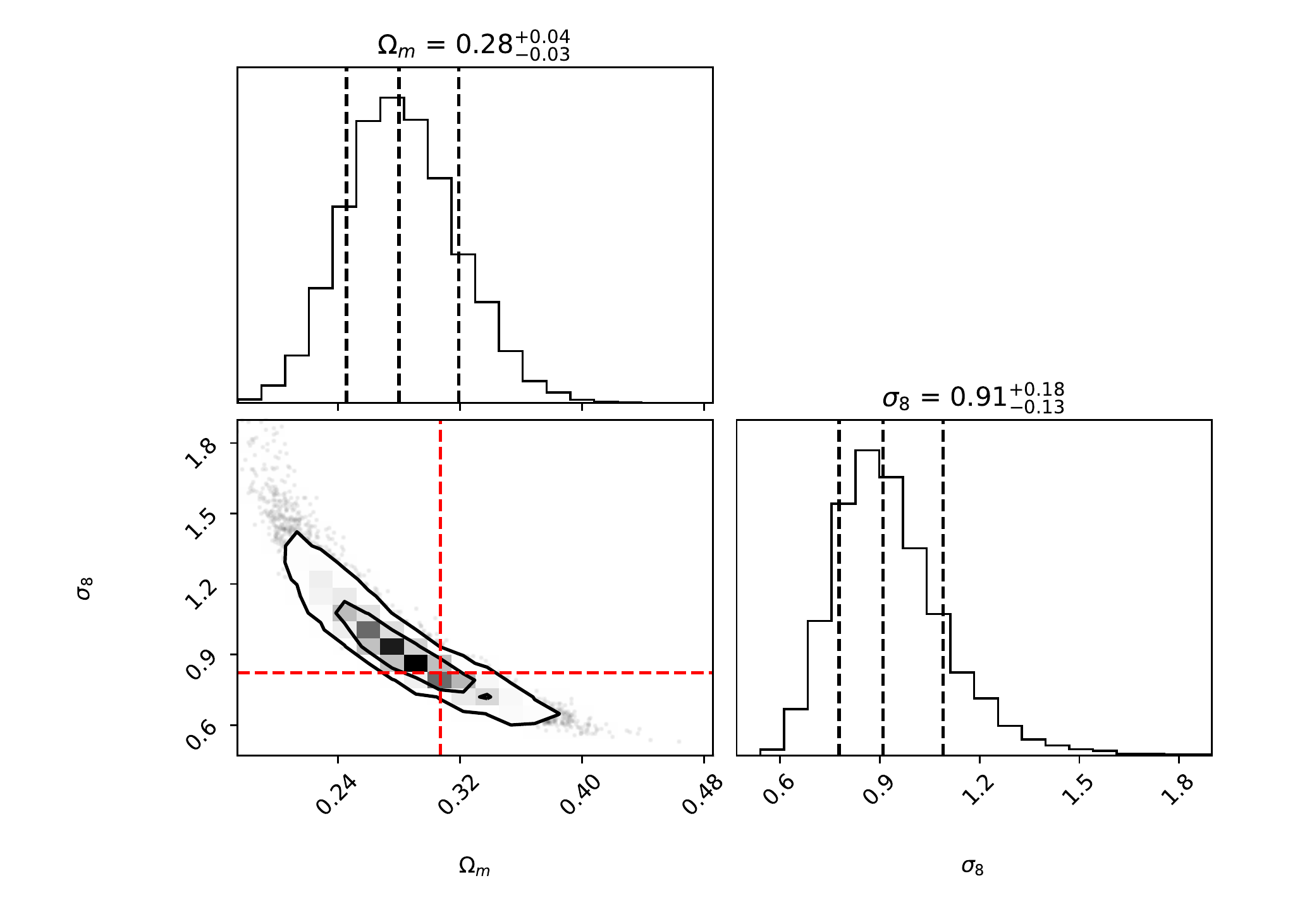}
   \includegraphics[width=\hsize]{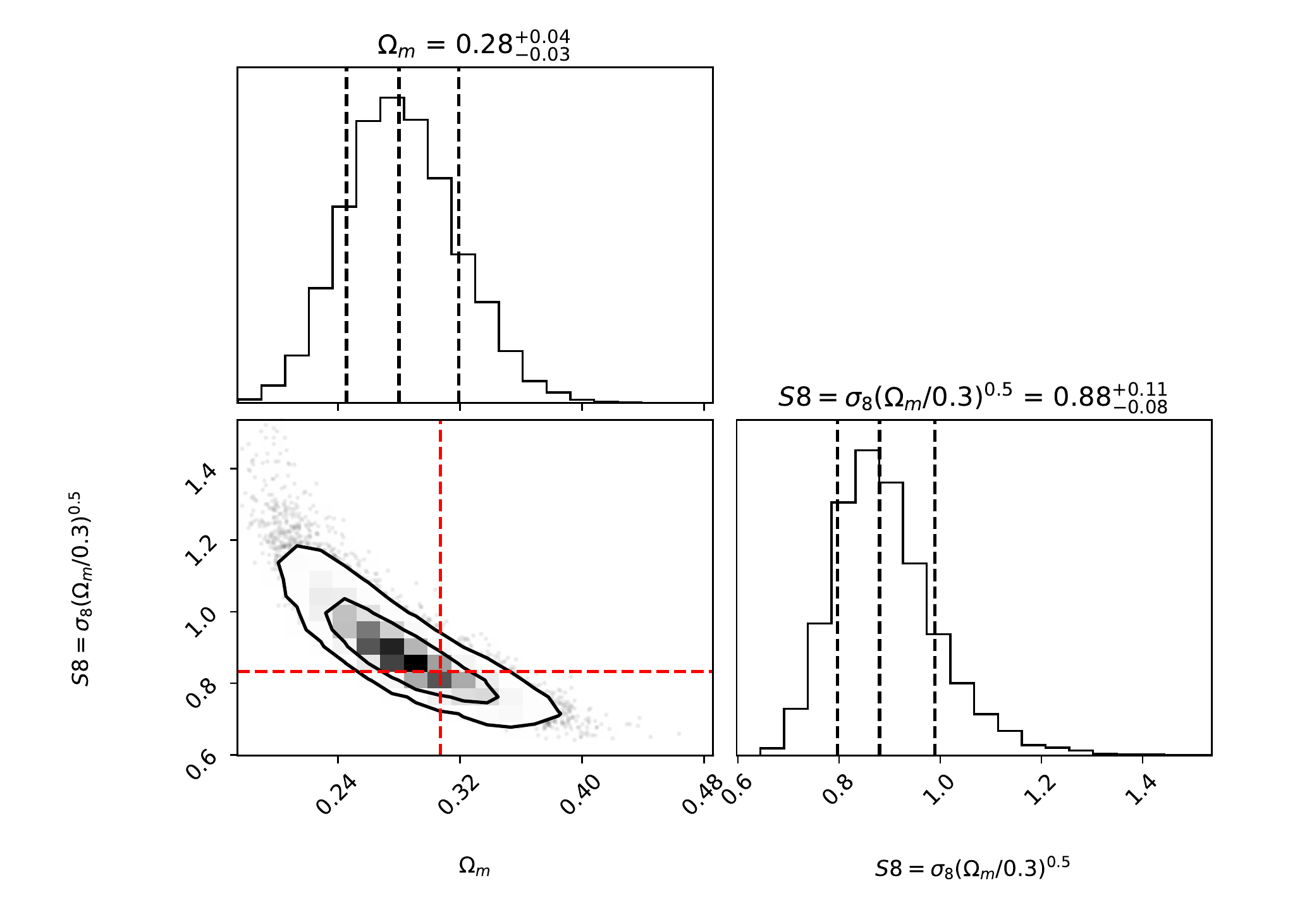}
      \caption{Posterior distribution for the parameters $\Om$ and $\sigma_8$ for the HMDPL catalog. The contours show the 68\% and 95\% confidence regions, and the values used in the simulation are shown by the dashed red lines. The dashed black lines in the marginalized posteriors show the 16\%, 50\%, and 84\% quantiles.}
         \label{fig:mdpl2_mcmc}
   \end{figure}

   To verify that we are able to recover the correct cosmology, we
   first ran the MCMC sampling using the HMDPL halo catalog. We used
   the same redshift, mass ranges, and binning for the projected
   two-point correlation function as in
   \sect{\ref{sec:validation}}. We also included photometric redshift
   errors estimated from the CODEX catalog. The resulting posterior
   distributions of the cosmological parameters are shown in Fig.
   ~\ref{fig:mdpl2_mcmc}, displaying the full two-dimensional likelihood
   contours and marginalized posterior distributions for $\Om$ and
   $\sigma_8$. The parameter values that were used to run the simulation are
   $\Om=0.307115$ and $\sigma_8=0.8228$. The marginalized constraints
   we obtain are $\Om = 0.28^{+0.04}_{-0.03}$ and $\sigma_8 =
   0.91^{+0.18}_{-0.13}$ , which are clearly compatible with the
   simulation cosmology.
   
\begin{figure}
   \centering
   \includegraphics[width=\hsize]{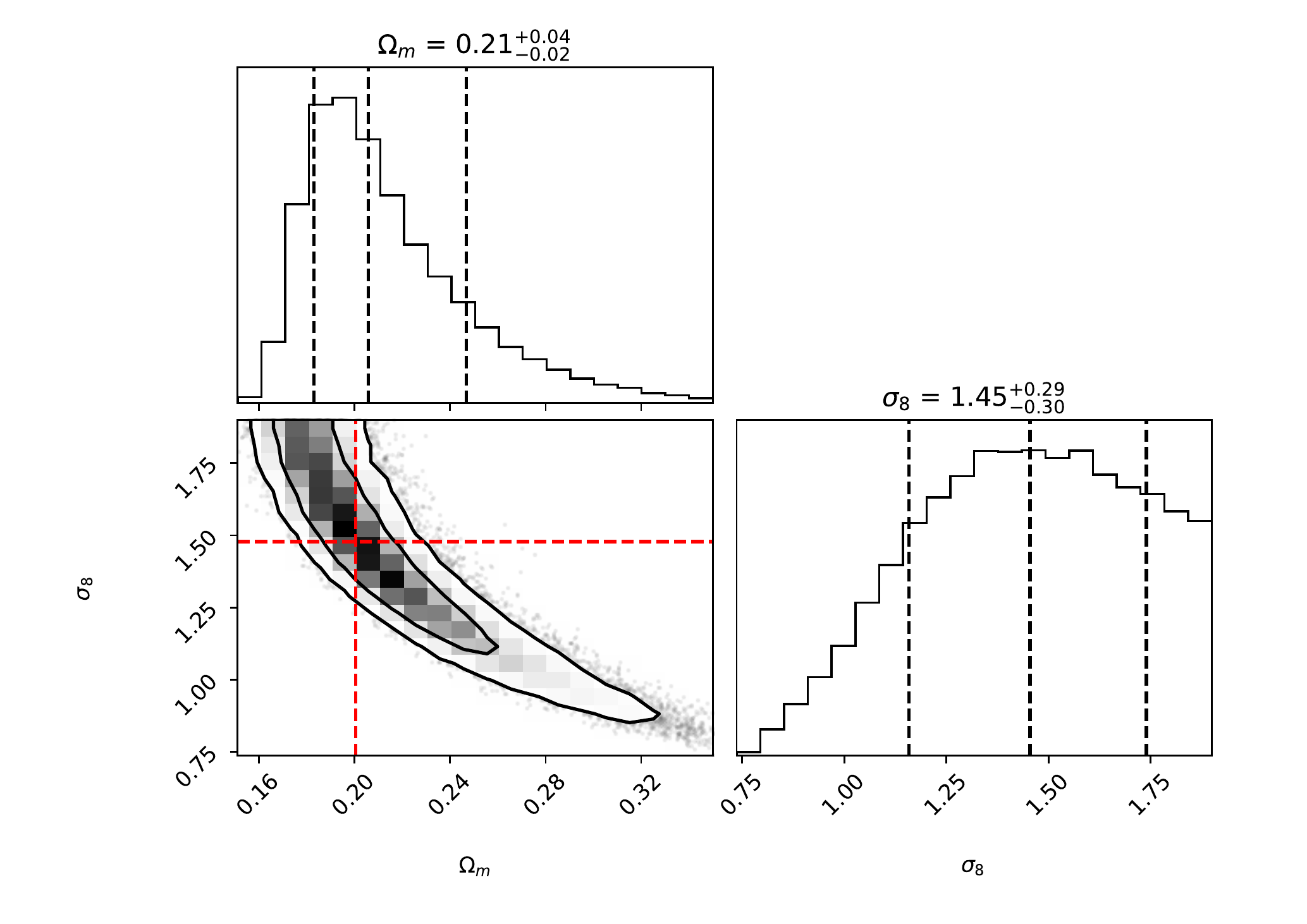}
      \caption{Posterior distribution for the parameters $\Om$ and $\sigma_8$ for the CODEX catalog in the full redshift range of $0.1 < z < 0.5$. The contours show the 68\% and 95\% confidence regions, and the best fit, values are shown by the dashed red lines. The dashed black lines in the marginalized posteriors show the 16\%, 50\%, and 84\% quantiles.}
         \label{fig:codex_mcmc_full}
   \end{figure}

\begin{figure}
   \centering
   \includegraphics[width=\hsize]{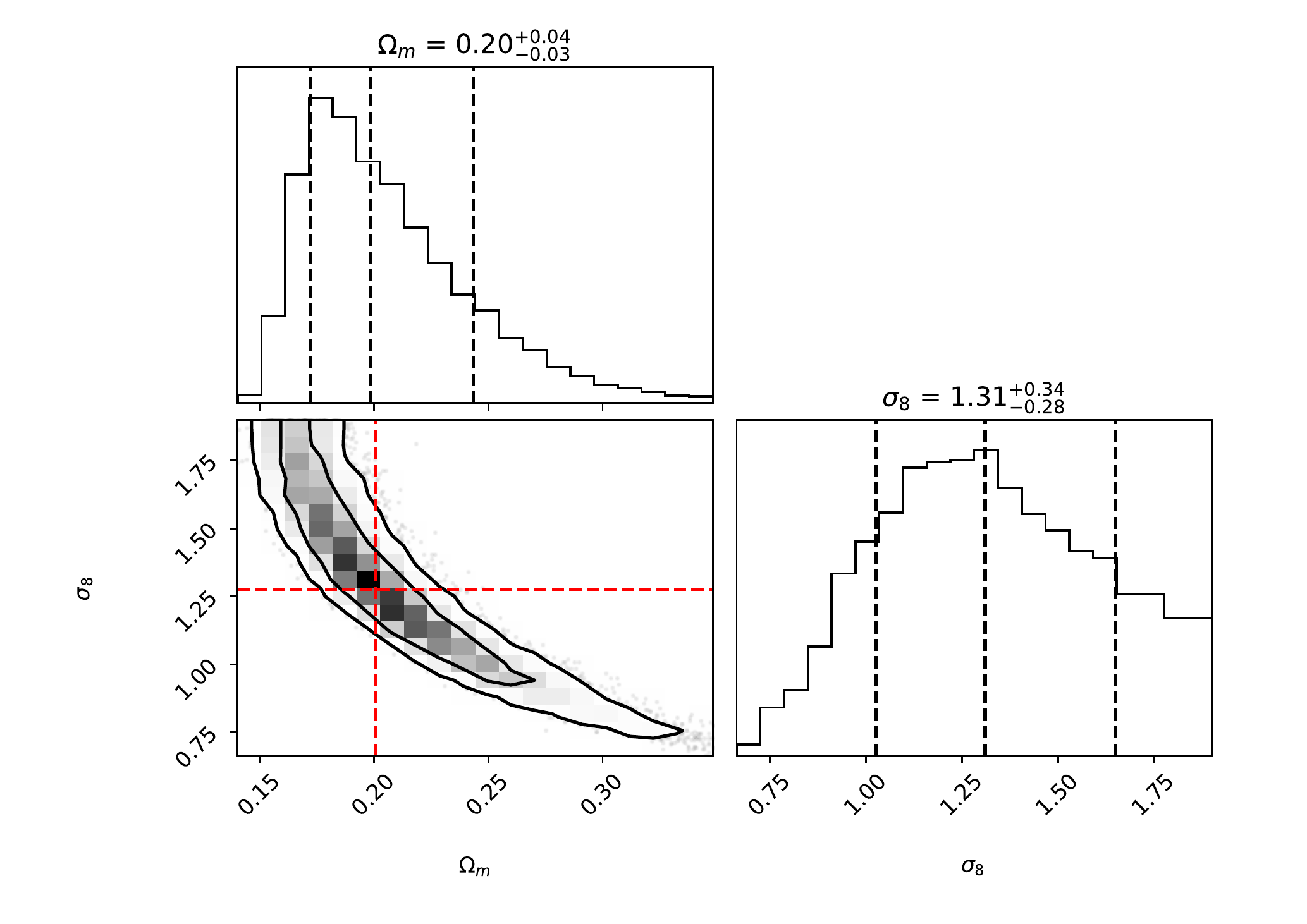}
      \caption{Same as \fig{\ref{fig:codex_mcmc_full},} but with the richness-to-mass scaling relation with $1\sigma$ deviation.}
         \label{fig:codex_mcmc_extreme}
   \end{figure}

   We then performed the same analysis on the CODEX clusters. All the
   details of the two-point correlation function estimates (such as
   binning) were exactly the same as described in
   \sect{\ref{sec:codex_clustering}}. We applied flat priors in the
   range of $0.05 < \Om < 0.5$ and $0.4 < \sigma_8 <
   1.9$. Figure~\ref{fig:codex_mcmc_full} shows the posterior
   distributions obtained for the full CODEX cluster catalog. It
   favors extremely high values of $\sigma_8$. The richness-to-mass
   scaling relation we used, however, has a quite significant
   uncertainty. When we take the lower $1\sigma$ extreme value for its
   normalization instead of the mean value, which is supported by
   calibrations obtained for Sunyaev-Zeldovich (SZ) clusters
   \citep{bleem}, we obtain the posterior distribution of
   \fig{\ref{fig:codex_mcmc_extreme}}. The obtained value of
   $\sigma_8$ is more compatible with canonical values. We adopt this
   $1\sigma$ lower limit as our baseline for the rest of the analysis.
   
\begin{figure}
   \centering
   \includegraphics[width=\hsize]{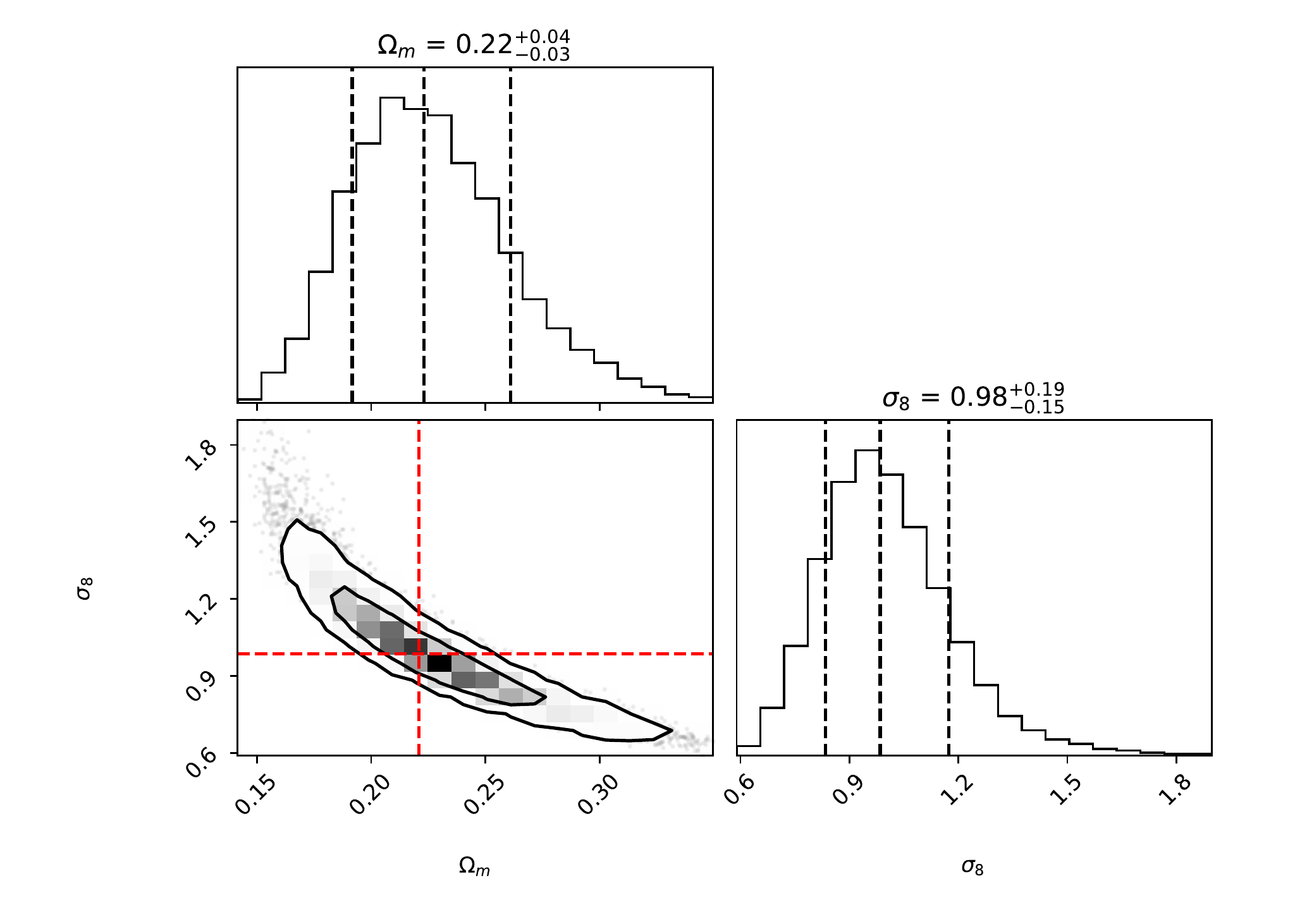}
   \includegraphics[width=\hsize]{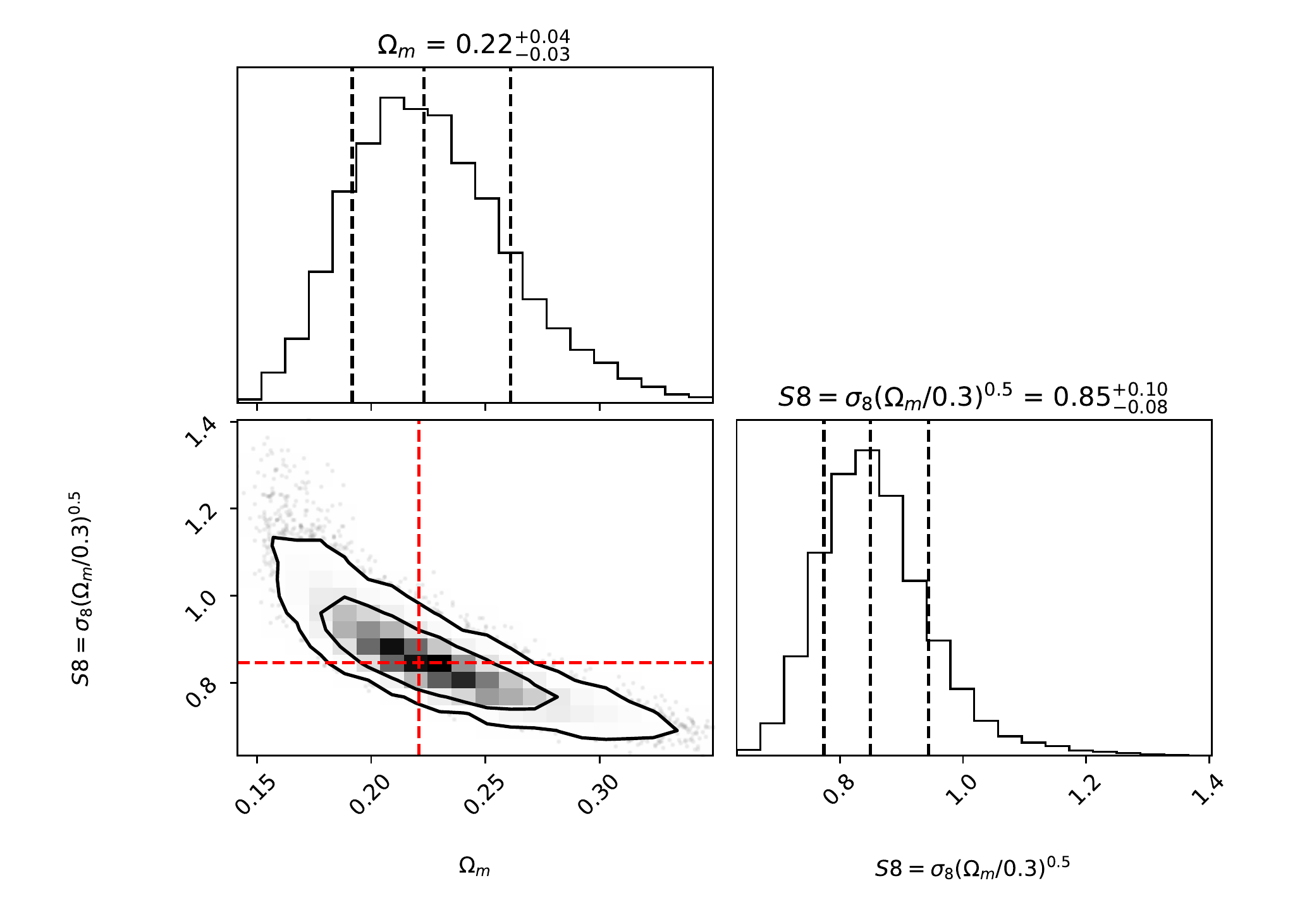}
      \caption{Posterior distribution for the parameters $\Om$, $\sigma_8$ , and $S_8$ for the CODEX catalog split into two redshift bins $0.1 < z < 0.3$ and $0.3 < z < 0.5$. The contours show the 68\% and 95\% confidence regions, and the best fit values are shown by the dashed red lines. The dashed black lines in the marginalized posteriors show the 16\%, 50\%, and 84\% quantiles.}
    \label{fig:codex_mcmc}
   \end{figure}

   For high values of $\sigma_8$ the distributions are very flat
   providing poor constraints. However, \cite{pillepich12} and
   \cite{pillepich18}, for example, showed that it is possible to make
   the angular clustering of clusters more sensitive to the cosmology
   by splitting the cluster sample into redshift bins. Motivated by
   this, we ran the MCMC sampling using two redshift bins, $0.1 < z <
   0.3$ and $0.3 < z < 0.5$, and computed a two-point correlation
   function estimate for both bins (see Fig.
   ~\ref{fig:codex_zcomparison} for a display). The resulting
   posterior distributions for $\Om$ and $\sigma_8$ are shown in
   \fig{\ref{fig:codex_mcmc}}. The distribution is in this case far
   less skewed toward high $\sigma_8$ , and we obtain marginalized
   parameter constraints of $\Om=0.22^{+0.04}_{-0.03}$ and
   $\sigma_8=0.98^{+0.19}_{-0.15}$ (or correcting for the dependency
   on $\Om$, $S_8=\sigma_8 (\Om
   /0.3)^{0.5}=0.85^{+0.10}_{-0.08}$). The constraints we obtain from
   the two redshift bins separately are $\Om = 0.22^{+0.05}_{-0.04}$,
   $\sigma_8 = 0.99^{+0.25}_{-0.20}$ for the low-redshift bin and $\Om
   = 0.22^{+0.08}_{-0.05}$, $\sigma_8 = 1.02^{+0.36}_{-0.24}$ for the
   high-redshift bin. When the two redshift bins are combined in the
   analysis, the constraints become tighter. However, the sampling was
   made assuming that the two redshifts bins are independent. This
   might not be the case, especially with photometric
   redshifts. Taking the covariance between the bins into account
   might loosen the constraints by some amount. For comparison with
   the WMAP9 cosmology, we also ran the MCMC sampling using
   \citet{planck2015} values for the fixed parameters. In this case,
   we obtain constraints $\Om = 0.23^{+0.03}_{-0.03}$ and $\sigma_8 =
   0.93^{+0.16}_{-0.13}$. Both are compatible with values using WMAP9
   cosmology within the statistical errors.

\begin{figure}
   \centering
   \includegraphics[width=\hsize]{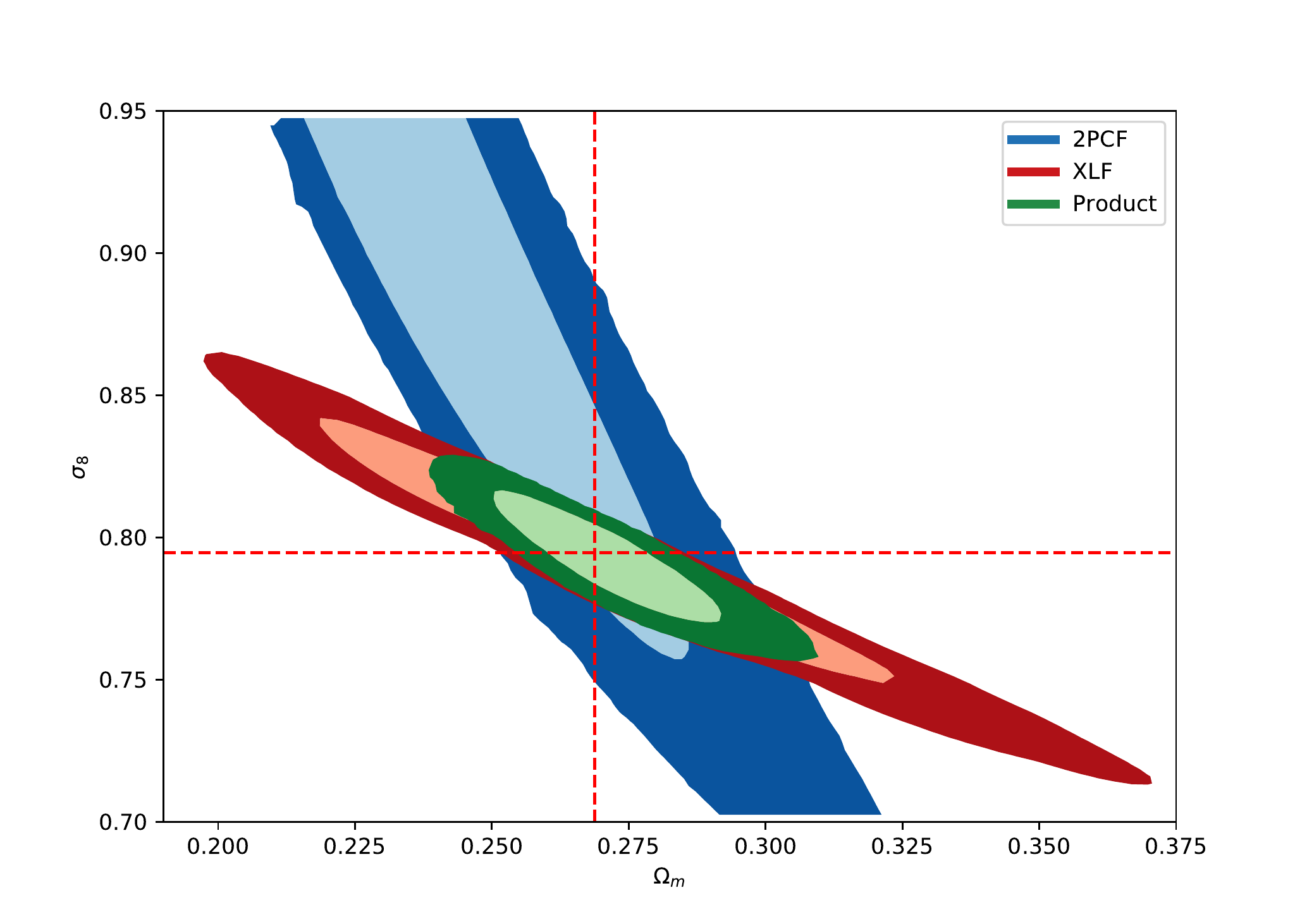}
   \includegraphics[width=0.92\hsize]{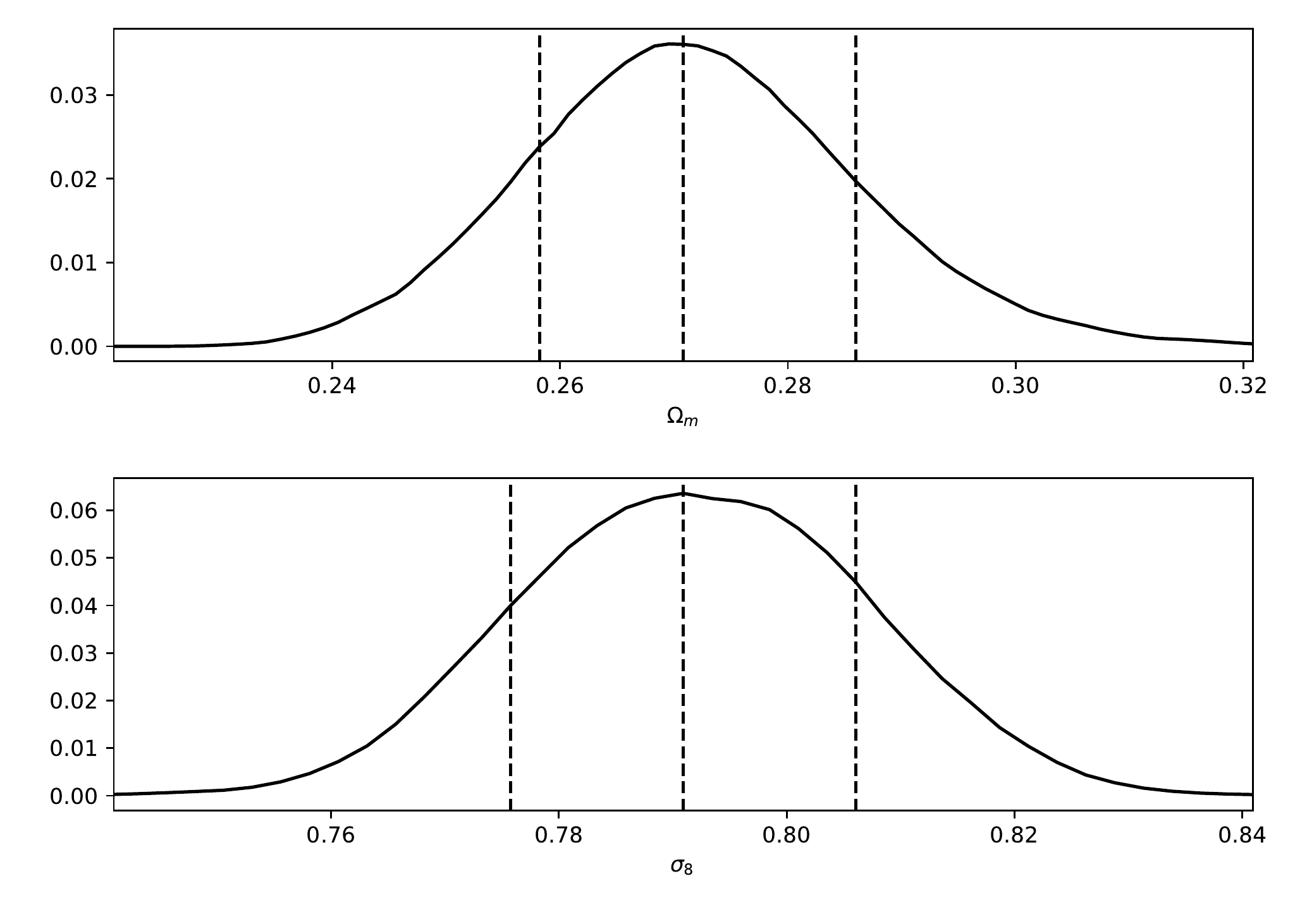}
      \caption{Parameter constraints combining two-point correlation function and luminosity function likelihoods. Top panel: Likelihood function of $(\Om$, $\sigma_8)$ from the cluster two-point correlation function (blue), X-ray luminosity function (red), and joint distribution (green). The light and dark contours are the 68 \% and 95 \% confidence regions, respectively, and the dashed red lines show the best-fit values for the joint likelihood. Bottom panel: Marginalized posterior distributions for $\Om$ and $\sigma_8$ obtained from the joint likelihood in the top panel. The dashed black lines show the 16\%, 50\%, and 84\% quantiles.}
         \label{fig:2pcfxxlf}
   \end{figure}

   \figs{\ref{fig:codex_mcmc_full}-\ref{fig:codex_mcmc}} show a strong
   degeneracy between $\Om$ and $\sigma_8$ when clustering data alone
   are used. It is possible to break some of this degeneracy by
   combining the likelihood from the clustering analysis with that
   from the cluster mass function. \citet{codex} used the cluster
   X-ray luminosity function, which is essentially a proxy for their
   mass function, to obtain constraints $\Om = 0.270 \pm 0.06$ and
   $\sigma_8 = 0.79 \pm 0.05$. In the top panel of
   \fig{\ref{fig:2pcfxxlf}} we show the constraints we obtain by
   combining the two likelihoods. The individual likelihood functions
   are mutually nearly orthogonal, and by combining them, we can
   significantly tighten the parameter constraints. The bottom panel
   of \fig{\ref{fig:2pcfxxlf}} shows the marginalized posterior
   distributions for $\Om$ and $\sigma_8$ obtained from the joint
   likelihood. From these we derive the parameter constraints $\Om =
   0.27^{+0.01}_{-0.02}$, $\sigma_8 = 0.79^{+0.02}_{-0.02}$. It should
   be noted, however, that we have estimated the joint likelihood by a
   simple product of the two likelihoods. This can result in an overly
   optimistic estimate because the two quantities involved are
   correlated; see, for example, \citet{lacasa16}, who estimated the
   cross-correlations of cluster counts (essentially their mass
   function) and galaxy power spectrum to be at $\sim 20 \%$ level.
   
     To estimate the effect of systematic uncertainties in our
     parameter constraints, we ran set of MCMC samplings using a variety of
     redshift measurements, selection functions, mass proxies, and
     bias-to-mass calibrations. A full account of this exercise is
     provided in Appendix~\ref{sec:systematics}. We estimate the
     systematic error of parameter $p$ as the sum of the squared
     differences between best-fit parameter values in the baseline
     case (presented in Fig.\ref{fig:codex_mcmc}) and the comparison
     cases,
     \begin{equation}
         \label{eq:systematics}
         \sigma_p = \sqrt{\frac{1}{N}\sum_{i=1}^N \left(\overline{p}-p_i\right)^2},
     \end{equation}
   where $\overline{p}$ is the best-fit value in the baseline case,
   and $p$ are the best fit values for the comparison cases. We
   separately estimated the systematic error originating from survey
   effects (redshift measurement and selection function) and the
   systematic error from predicting the bias. The results along with
   the combination of the two error categories are shown in
   \tb{\ref{tab:systematics}}. The total systematic errors are
   $\sigma_{\Om} = 0.02$ and $\sigma_{\sigma_8} = 0.23$ for the
   two-point correlation function alone and $\sigma_{\Om} =0.07 $ and
   $\sigma_{\sigma_8} = 0.04$ for the combination of the two-point
   correlation function and the X-ray luminosity function. Both are
   of similar magnitude than the statistical errors in their respective cases.
    \begin{table}
        \caption{Systematic errors from survey effects and bias prediction, and their combination. The second and third column list the two-point correlation function alone, and the fourth and fifth column list the combination of the two-point correlation function and the X-ray luminosity function.}
        \label{tab:systematics}
        \centering
            \begin{tabular}{c c c c c}
            \hline\hline
            & \multicolumn{2}{c}{2PCF} & \multicolumn{2}{c}{2PCF $\times$ XLF}\\
            Source & $\Om$ & $\sigma_8$ & $\Om$ & $\sigma_8$ \\
            \hline
            Survey & 0.018 & 0.031 & 0.014 & 0.017\\
            Mass & 0.005 & 0.231 & 0.070 & 0.039\\
            Combined & 0.019 & 0.233 & 0.072 & 0.042\\
            \hline
        \end{tabular}
    \end{table}
    
   A summary of our results along with a comparison with the WMAP9 and
   Planck 2018 results \citep{p18} is listed in \tb{\ref{tab:cosmology}}. Our
   results from combining clustering and mass function are fully
   consistent with the WMAP9 results within the statistical
   uncertainty. A difference larger than $1\sigma$ between our values
   and those from Planck 2018 for $\Om$ is visible, but this
   difference is smaller than our estimated systematic uncertainty.
    \begin{table}
        \caption{Summary of cosmological constraints obtained from the cluster two-point correlation function, two-point correlation function combined with the X-ray luminosity function, the X-ray luminosity function alone, and for CMB datasets. Our error estimates include only the statistical errors.}
        \label{tab:cosmology}
        \centering
            \begin{tabular}{c c c}
            \hline\hline
            Dataset & $\Om$ & $\sigma_8$\\
            \hline
            2PCF & $0.22^{+0.04}_{-0.03}$ & $0.98^{+0.19}_{-0.15}$\\
            2PCF $\times$ XLF & $0.27^{+0.01}_{-0.02}$ & $0.79^{+0.02}_{-0.02}$ \\
            CODEX XLF & $0.27 \pm 0.06$ & $0.79 \pm 0.05$\\
            WMAP9 & $0.279 \pm 0.025$ & $0.821 \pm 0.023$ \\
            Planck 2018 & $0.3147 \pm 0.0074$ & $0.8101 \pm 0.0061$ \\
            \hline
        \end{tabular}
    \end{table}
   All the posterior distribution corner plots in this work were plotted using the python library corner\footnote{\url+https://github.com/dfm/corner.py+} \citep{corner}.
   
\section{Conclusions}
   We performed a clustering analysis on the CODEX galaxy cluster
   catalog. As a part of this analysis, we aimed at predicting the
   clustering bias of these clusters based on their masses. We first
   verified with a halo catalog from the HMDPL simulation that using
   this approach we can predict the clustering bias of dark matter
   halos with perfectly known masses in a known cosmology. We showed
   that we can recover the observed clustering bias within the
   statistical errors by simply averaging over the mass-based bias
   prediction for each individual halo in the sample. All three
   components in the analysis, the clustering of galaxy clusters (or
   dark matter halos in the case of simulations), the expected dark
   matter distribution, and the mass-to-bias conversion, depend on the
   cosmology, thus the agreement of them is a test of the cosmological
   model. We performed an MCMC sampling of parameters $\Om$ and
   $\sigma_8$ to find the best-fit values by comparing the measured
   two-correlation function of HMDPL halos to the total matter
   distribution scaled by the predicted bias factor. We recovered the
   input cosmological parameters within the statistical uncertainty.
   
   We applied the same analysis to the CODEX catalog. Cluster masses
   can be estimated from their X-ray luminosities or their
   richness. We determined which of these estimates of bias was best
   compatible with the data. We found that the mass estimates predict
   bias factors that agree with the measured value at a level of
   17-39\%; the richness-based estimates give the best agreement. We
   also tested how splitting the sample into redshifts bins affects
   the results and found that the predicted bias agrees with the
   measured value at the 5-17\% level depending on the redshift range
   used.  We applied the same MCMC sampling as to the HMDPL halo
   catalog to CODEX clusters. We found that in order to have a
   constraining power on $\sigma_8$ , we need to split the cluster
   sample into redshift bins (two in the case of our analysis). By
   binning the sample according to redshift, we obtained the following
   parameter constraints: $\Om=0.22^{+0.04}_{-0.03}$ and
   $\sigma_8=0.98^{+0.19}_{-0.15}$. We estimated an additional error
   of $\pm 0.02$ and $\pm 0.19$, respectively, that originates from
   systematic effects related to survey effects and bias
   modeling. After combining the clustering-based likelihood with one
   from the cluster mass function, we obtained the following parameter
   constraints: $\Om = 0.27^{+0.01}_{-0.02}$ and $\sigma_8 =
   0.79^{+0.02}_{-0.02}$. In this case, we estimated that the
   systematic uncertainties contribute $\pm 0.07$ and $\pm 0.04$
   additionally, respectively. It should be noted, however, that
   proper handling of the covariance between the two quantities would
   most likely loosen the constraints by some amount. In any case, our
   parameter constraints from clustering bias are consistent with the
   WMAP nine-year cosmology, and when systematic uncertainties are
   included, also with the Planck 2018 cosmology.

\begin{acknowledgements}
The authors wish to acknowledge CSC -- IT Center for Science, Finland, for computational resources.
 
We acknowledge grants of computer capacity from the Finnish Grid and Cloud Infrastructure (persistent identifier urn:nbn:fi:research-infras-2016072533 ).

Based on observations made with the Nordic Optical Telescope, operated by the Nordic Optical Telescope Scientific Association at the Observatorio del Roque de los Muchachos, La Palma, Spain, of the Instituto de Astrofisica de Canarias.
 
 The authors gratefully acknowledge the Gauss Centre for Supercomputing e.V. (www.gauss-centre.eu) and the Partnership for Advanced Supercomputing in Europe (PRACE, www.prace-ri.eu) for funding the MultiDark simulation project by providing computing time on the GCS Supercomputer SuperMUC at Leibniz Supercomputing Centre (LRZ, www.lrz.de).
 
 Funding for the Sloan Digital Sky Survey IV has been provided by the Alfred P. Sloan Foundation, the U.S. Department of Energy Office of Science, and the Participating Institutions. SDSS-IV acknowledges support and resources from the Center for High-Performance Computing at
the University of Utah. The SDSS web site is www.sdss.org.

SDSS-IV is managed by the Astrophysical Research Consortium for the Participating Institutions of the SDSS Collaboration including the Brazilian Participation Group, the Carnegie Institution for Science, Carnegie Mellon University, the Chilean Participation Group, the French Participation Group, Harvard-Smithsonian Center for Astrophysics, Instituto de Astrof\'isica de Canarias, The Johns Hopkins University, Kavli Institute for the Physics and Mathematics of the Universe (IPMU) / University of Tokyo, the Korean Participation Group, Lawrence Berkeley National Laboratory, Leibniz Institut f\"ur Astrophysik Potsdam (AIP), Max-Planck-Institut f\"ur Astronomie (MPIA Heidelberg), Max-Planck-Institut f\"ur Astrophysik (MPA Garching), Max-Planck-Institut f\"ur Extraterrestrische Physik (MPE), National Astronomical Observatories of China, New Mexico State University, New York University, University of Notre Dame, Observat\'ario Nacional / MCTI, The Ohio State University, Pennsylvania State University, Shanghai Astronomical Observatory, United Kingdom Participation Group, Universidad Nacional Aut\'onoma de M\'exico, University of Arizona, University of Colorado Boulder, University of Oxford, University of Portsmouth, University of Utah, University of Virginia, University of Washington, University of Wisconsin, Vanderbilt University, and Yale University.
\end{acknowledgements}


\appendix
\section{Systematic effects}
\label{sec:systematics}
   We report the different systematic effects in the cosmological
   parameter constraints we obtain. Our baseline case is the one where
   we split the CODEX sample in two redshift bins $0.1 < z < 0.3$ and
   $0.3 < z < 0.5$ and use richness-based mass estimates and the
   $b(M)$ model from \citet{comparat17}. We compared the best-fit
   parameter values for $\Om$ and $\sigma_8$ from MCMC samplings by
   varying survey effects, mass estimates, and the $b(M)$ model. All
   of the best-fit parameter values we obtain for different comparison
   cases are summarized in \tb{\ref{tab:parameter_comparison}}.

\begin{figure}
   \centering
   \includegraphics[width=\hsize]{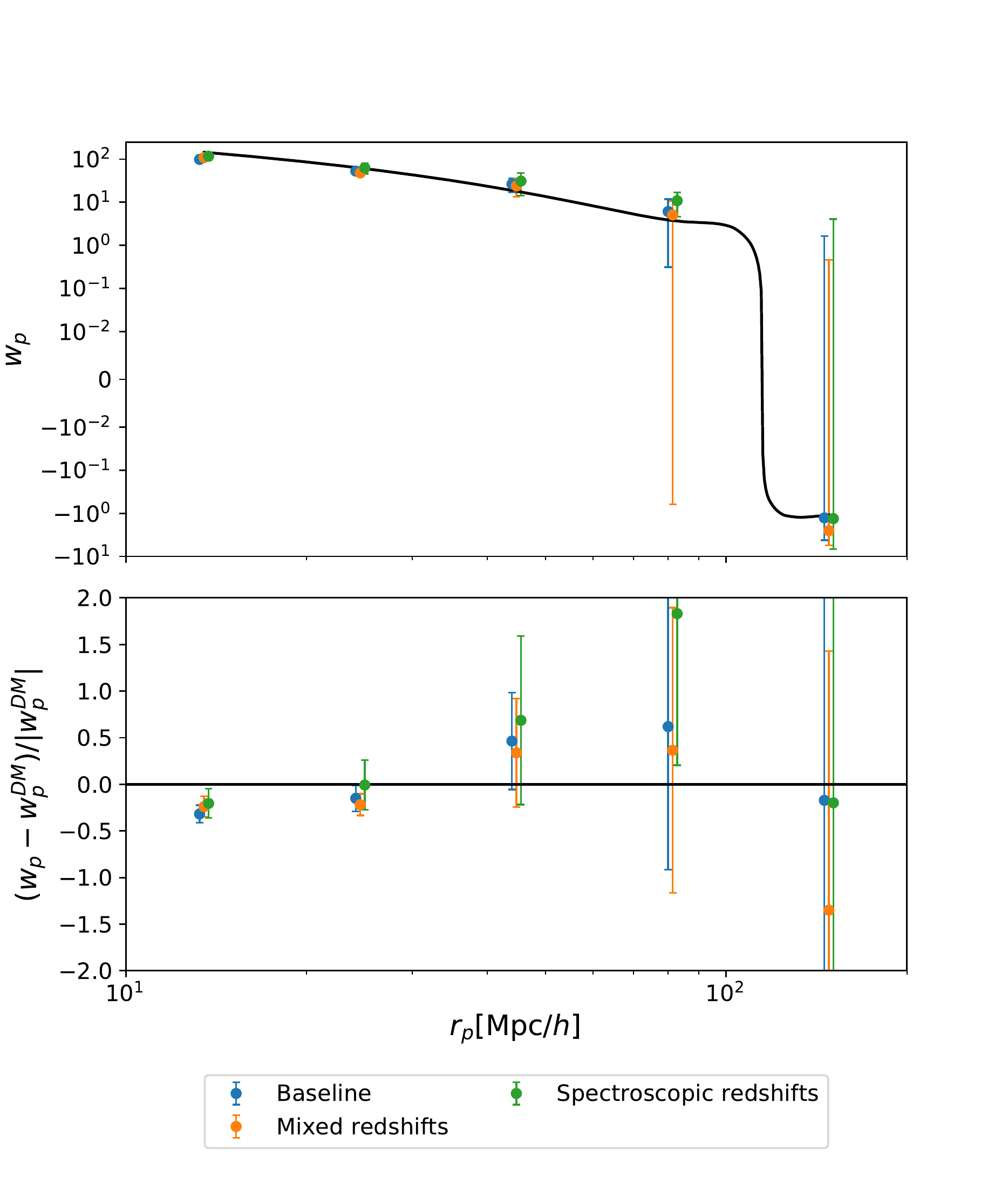}
      \caption{Comparison of photometric, mixed, and spectroscopic redshifts. Top panel: Projected two-point correlation functions in the redshift range $0.1 < z < 0.5$. Data points are the two-point correlation function estimate from CODEX clusters. The solid curve is the predicted dark matter two-point correlation function scaled by $\overline{b}^{\; 2}$ computed from the halo mass estimates. Bottom panel: Relative difference between the measured and predicted two-point correlation functions. The data points have been shifted horizontally for clarity.}
         \label{fig:codex_zcomparison_mix}
   \end{figure}

   As mentioned in Sec.~\ref{sec:validation}, a subset of CODEX
   clusters has spectroscopic redshifts associated with them in
   addition to photometric redshifts. To verify the robustness of the
   photometric redshifts, we computed the two-point correlation
   function using spectroscopic redshifts when available and
   photometric redshifts for the remaining clusters. A comparison
   between measured and predicted correlation functions in the full
   redshift range is presented in
   \fig{\ref{fig:codex_zcomparison_mix}}, where we also show the
   results for a purely spectroscopic sample limited to the SDSS DR16
   area. In this case, we have a slightly smaller set of spectroscopic
   redshifts (880 within $0.1 < z < 0.5$) because we required
   spectroscopic completeness. The measured and predicted biases for
   both cases in all the redshift bins are listed in
   \tb{\ref{tab:bias_zcomparison_mix}}.  At lower redshifts, the
   measured biases are compatible with those from the purely
   photometric sample, but at higher redshifts, there is some
   difference.
   \begin{table*}
        \caption{Bias values obtained for a mixture of photometric and spectroscopic redshifts (column $b_{\mathrm{mix}}$) and those of a purely photometric sample (column $b_{\mathrm{phz}}$), and a purely spectroscopic sample (columns $b_{\mathrm{spe}}$) as well as the mass-based prediction (column $\overline{b}$) from \tb{\ref{tab:bias_zcomparison}}. Column $n_{\mathrm{spe}}$ shows the number of clusters with spectroscopic redshifts in each redshift bin.}
        \label{tab:bias_zcomparison_mix}
        \centering
        \centerline{
            \begin{tabular}{c c c c c c}
            \hline\hline
            Redshift range &  $n_{\mathrm{spe}}$ & $b_{\mathrm{phz}}$ & $b_{\mathrm{spe}}$ & $b_{\mathrm{mix}}$ & $\overline{b}$\\
            \hline
            $0.1 < z < 0.5$ & 1223 & $3.70 \pm 0.13$ & $4.07 \pm 0.30$ & $3.77 \pm 0.08$ & 4.33\\
            $0.1 < z < 0.3$ & 823 & $3.78 \pm 0.10$ & $3.90 \pm 0.27$ & $3.72 \pm 0.09$ & 3.95\\
            $0.3 < z < 0.5$ & 400 & $4.71 \pm 0.66$ & $6.67 \pm 0.40$ & $6.45 \pm 0.77$ & 5.08\\
            \hline
        \end{tabular}
        }
    \end{table*}

\begin{figure}
        \centering
        \includegraphics[width=\hsize]{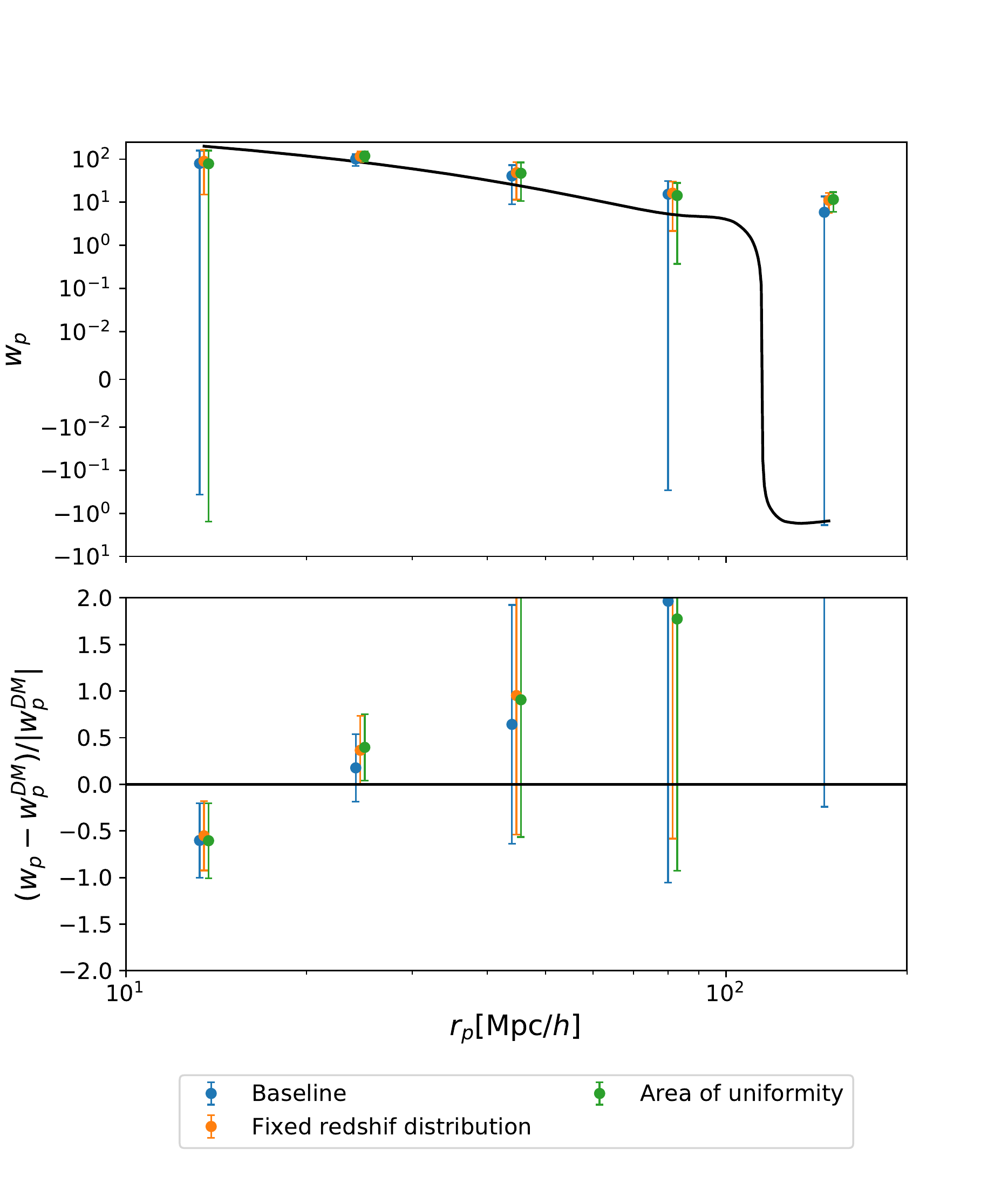}
        \caption{Comparison of different selection functions in high-z bin. Top panel: Projected two-point correlation function in redshift range $0.3 < z < 0.5$. Data points are the two-point correlation function estimate from CODEX clusters. The solid curve shows the predicted dark matter two-point correlation function scaled by $\overline{b}^{\; 2}$ computed from the halo mass estimates. Bottom panel: Relative difference between the measured and predicted two-point correlation functions. The data points have been shifted horizontally for clarity.}
        \label{fig:selection}
    \end{figure}

    In \fig{\ref{fig:codex_zcomparison}} the slope of the measured
    correlation function in the high-redshift bin and at large scales
    appears to be flatter than the prediction and the low-redshift
    bin. To test whether this is due to an incorrect characterization
    of the selection function, we ran the analysis forcing the
    redshift distribution of the random catalog to be exactly that of
    the data catalog in redshift bins of $\Delta z = 0.027$. Another
    test we ran was to use the subregion of the CODEX survey area in
    which the X-ray sensitivity is uniform. A comparison of the fitted
    biases obtained using different selection functions is given in
    \tb{\ref{tab:bias_selection}}.
    \begin{table}
        \caption{Comparison of fitted  and predicted biases using different selection functions. Column $b$ is the fitted bias, $\overline{b}$ is the mass-based prediction. "Full" indicates the whole CODEX survey area, and "Uniform" the subregion of uniform X-ray sensitivity. Column $n_D$ shows the number of clusters in each combination of redshift range and sky area.}
        \label{tab:bias_selection}
        \centering
            \begin{tabular}{c c c c c}
            \hline\hline
            Redshift range & Sky area & $n_D$ & $\overline{b}$ & $b$ \\
            \hline
            $0.1 < z < 0.5$ & Full & 1892 & 4.33 & $3.91 \pm 0.14$ \\
            $0.1 < z < 0.3$ & Full & 1250 & 3.95 & $3.68 \pm 0.18$ \\
            $0.3 < z < 0.5$ & Full & 642  & 5.08 & $6.64 \pm 2.17$ \\
            \hline
            $0.1 < z < 0.5$ & Uniform & 1831 & 4.32 & $3.92 \pm 0.16$ \\
            $0.1 < z < 0.3$ & Uniform & 1206 & 3.94 & $3.66 \pm 0.21$ \\
            $0.3 < z < 0.5$ & Uniform & 625  & 5.06 & $7.02 \pm 1.49$ \\
            \hline
        \end{tabular}
    \end{table}
   \fig{\ref{fig:selection}} shows the two-point correlation function results in a high-z bin using different selection functions. They show that the flattening of the measured two-point correlation function at large scales at high redshifts occurs regardless of the selection function used, and we therefore suspect that this is not the cause of the effect.
The effect of redshift errors and selection function on the cosmological parameter constraints is demonstrated by the corresponding changes in the best-fit parameter values presented in \tb{\ref{tab:parameter_comparison}}.

  In addition to different survey effects, we studied the effect of
  varying the mass and bias estimates. We ran MCMC sampling using the
  two X-ray luminosity based mass calibrations (introduced in
  \sect{\ref{sec:codex_clustering}}). From the bias models $b(M)$ in
  \tb{\ref{tab:biascomparison}} we chose the two that gave the most
  extreme values, that is, \citet{tinker10} and \citet{bhattacharya}, 
  to be compared with the baseline case. The
  resulting shifts in the best-fit parameter values are shown again in
  \tb{\ref{tab:parameter_comparison}}.
   \begin{table}
        \caption{Best-fit parameter values for all the cases used to compute the systematic error on the parameters $\Om$ and $\sigma_8$. The second and third column show clustering data alone, and the fourth and fifth column list the combination of clustering and X-ray luminosity function.}
        \centering
        \centerline{
        \label{tab:parameter_comparison}
            \begin{tabular}{c c c c c}
            \hline\hline
            & \multicolumn{2}{c}{2PCF} & \multicolumn{2}{c}{2PCF $\times$ XLF}\\
            Case & $\Om$ & $\sigma_8$ & $\Om$ & $\sigma_8$\\
            \hline
            Baseline & 0.22 & 0.99 & 0.27 & 0.79\\
            Fixed $dn/dz$ & 0.21 & 1.04 & 0.26 & 0.80\\
            Fixed $dn/dz$, area of uniformity & 0.22 & 0.97 & 0.25 & 0.81\\
            Mixed redshifts & 0.19 & 1.10 & 0.26 & 0.81\\
            X-ray mass, \citet{codex} & 0.22 & 1.35 & 0.37 & 0.78\\
            X-ray mass, \citet{capasso20} & 0.22 & 1.17 & 0.33 & 0.74\\
            $b(M)$, \citet{tinker10} & 0.23 & 1.10 & 0.32 & 0.76\\
            $b(M)$, \citet{bhattacharya} & 0.22 & 0.79 & 0.21 & 0.84 \\
            \hline
        \end{tabular}
        }
    \end{table}
\end{document}